\documentclass{aa}  

\usepackage{graphicx}   
\usepackage{amsmath}	
\usepackage{txfonts}
\newsavebox{\measurebox}
\usepackage{subcaption}
\usepackage{color}
\usepackage{ulem}
\usepackage{adjustbox}
\usepackage{tablefootnote}
\usepackage{threeparttable}
\usepackage[justification=raggedright]{caption}
\usepackage{booktabs}
\usepackage{tabularx}
\usepackage{soul}
\usepackage{gensymb}
\usepackage[dvipsnames]{xcolor}
\usepackage{multicol}
\usepackage{multirow}
\usepackage{placeins}
\usepackage{hyperref}
\begin{document}

   \title{MeerKAT discovery of GHz radio emission extending from Abell~3017 toward Abell~3016}

   \subtitle{}
   \titlerunning{Diffuse radio emission within A3017-A3016}
   \authorrunning{Hu D., et al.}

   \author{Dan Hu \inst{1},
           Norbert Werner \inst{1},
           Haiguang Xu \inst{2},
           Qian Zheng \inst{3},
           Jean-Paul Breuer \inst{4,1},
           Linhui Wu \inst{3},
           Stefan~W. Duchesne \inst{5}, 
           Reinout~J. van Weeren \inst{6},
           Ming Sun \inst{7},
           Congyao Zhang \inst{1},
           Melanie Johnston-Hollitt \inst{8},
           Huanyuan Shan \inst{3},
           Quan Guo \inst{3},
           Zhenghao Zhu \inst{3},
           Jingying Wang \inst{3},
           Junhua Gu \inst{9},
           Yuanyuan Zhao \inst{2},
           Hoongwah Siew \inst{2},
           Junjie Mao \inst{10},
           Zhongli Zhang \inst{3,11},
           and 
           Tomáš Plšek \inst{1}
    }

   \institute{
        Department of Theoretical Physics and Astrophysics, Faculty of Science, Masaryk University, Kotl\'{a}\v{r}sk\'{a} 2, Brno 611 37, Czech Republic\\
        \email{hudan.bazhaoyu@mail.muni.cz}
        \and
        School of Physics and Astronomy, Shanghai Jiao Tong University, Dongchuan Road 800, Shanghai 200240, China 
        \and
        Shanghai Astronomical Observatory, Chinese Academy of Sciences, Nandan Road 80, Shanghai, China
        \and
        Department of Physics, Graduate School of Advanced Science and Engineering, Hiroshima University Kagamiyama, 1-3-1 Higashi-Hiroshima, 739-8526, Japan 
        \and
        CSIRO Space \& Astronomy, PO Box 1130, Bentley, WA 6102, Australia
        \and
        Leiden Observatory, Leiden University, PO Box 9513, 2300 RA Leiden, The Netherlands
        \and
        Department of Physics and Astronomy, University of Alabama in Huntsville, Huntsville, AL 35899, USA
        \and
        Curtin Institute for Computation, Curtin University, GPO Box U1987, Perth 6845, WA, Australia
        \and
        National Astronomical Observatories, Chinese Academy of Sciences, 20A Datun Road, Beĳing 100101, P. R. China
        \and
        Department of Astronomy, Tsinghua University, Beijing 100084, People's Republic of China
        \and
        Key Laboratory of Radio Astronomy and Technology, Chinese Academy of Sciences, A20 Datun Road, Chaoyang District, Beijing, 100101, P. R. China
        }

   \date{Received -- , ; accepted -- , }

 
  \abstract
   {Cosmic filaments are vast, faint structures that connect galaxy clusters, often challenging to detect directly. However, filaments between pre-merger cluster pairs become more visible due to gas heating and compression while the clusters are approaching, enabling detection in X-ray and radio wavelengths.
   The clusters Abell~3017 and Abell~3016 are located within such a large-scale filament. A prominent X-ray bridge has been detected connecting the two clusters and a potential galaxy group between them.}
   {The aim of this work is to investigate the existence of a radio bridge in the filament between Abell~3017 and Abell~3016, to explore other diffuse radio structures within this system, and to investigate the origins of these diffuse radio emission.}
   {We analyzed MeerKAT L-band data to study the morphology and spectra of the diffuse radio structures in Abell~3016-Abell~3017. X-ray imaging and spectral analysis were carried out with archival Chandra and XMM-Newton data. Additionally, correlations between radio ($I_R$) and X-ray surface brightness ($I_X$) were generated to explore the connections between thermal and non-thermal components in the diffuse radio emission.}
   {We detected a faint radio bridge with an average surface brightness of $\sim 0.1~\mu\rm Jy~arcsec^{-2}$ at 1280 MHz using MeerKAT. It connects Abell~3017 with a potential galaxy group and extends towards Abell~3016, aligning with the X-ray bridge. A high X-ray temperature of $7.09 \pm 0.54$~keV detected in the bridge region suggests an interaction between Abell~3017 and the group.
   In Abell~3017, we identified two distinct components of diffuse radio emission: a radio mini-halo and an outer radio halo with a northern extension (N-extension hereafter). The radio surface brightness profile of Abell~3017 shows a steep inner component consistent with other mini-halos, and a faint outer component likely linked to an infalling subcluster. The $I_{\rm R}-I_{\rm X}$ diagram indicates superlinear and sublinear correlations for the mini-halo and N-extension, respectively.}  
   {We proposed three plausible explanations for the origin of the radio bridge: (1) it is an inter-cluster radio bridge connecting the two clusters in a filament, enhanced by interactions with the embedded galaxy group; (2) it results from an interaction between Abell~3017 and the galaxy group after their primary apocentric passage, with the group currently falling back towards Abell~3017; (3) it is a cluster radio relic associated with a merger shock, appearing as a bridge due to its face-on orientation. 
   In Abell~3017, the mini-halo is likely powered by gas sloshing, resulting from an offset merger that left the cluster's cool core intact. Turbulence from an infalling subcluster likely contributes to the formation of the outer radio halo.}

   \keywords{galaxies: clusters: general -- galaxies: clusters: individual: Abell~3017 -- galaxies: clusters: individual: Abell~3016 -- radio continuum: general -- X-rays: galaxies: clusters}

   \maketitle
%

\section{Introduction}

The cosmic structure of the Universe is characterized by a vast web of interconnected components, including galaxy clusters and cosmic filaments. Galaxy clusters, which are the most massive gravitationally bound systems, form primarily through the accumulation of matter along these filaments, where dark matter and gas converge. As matter flows through the filaments, it becomes gravitationally bound, leading to the creation of new clusters or the growth of existing ones. The dynamics of these structures are further influenced by cluster mergers, which occur when two or more clusters collide and combine, often facilitated by their proximity to cosmic filaments. These merger events are among the most energetic phenomena in the universe and play a critical role in shaping the properties and evolution of clusters. 
Galaxy clusters provide a unique environment for studying astrophysical processes across a wide range of scales, from sub-kiloparsecs (kpc) to megaparsecs (Mpc), including phenomena such as merger-induced gas sloshing, bulk motion, shock front, and active galactic nucleus (AGN) feedback. Conversely, cosmic filaments are generally faint structures that become more visible between pre-merger clusters due to the heating and compression of gas caused by preceding shock waves and turbulence. This process enhances their visibility, making them detectable in both X-ray and radio wavelengths.

In recent decades, advances in radio telescope sensitivity and resolution, combined with improved data analysis techniques, have led to the discovery and investigation of an increasing number of radio sources within galaxy clusters.
Diffuse radio emission, a form of radio source characterized by its extended nature and not necessarily associated directly with individual cluster radio galaxies, is often detected in cluster environments and can encompass several types, such as radio halos, mini-halos, and relics (see reviews by \citealt{feretti12,BJ14,vanweeren19}).

Radio halos, commonly found in massive and dynamically active galaxy clusters, are centrally located and coincide with X-ray emission from the intracluster medium (ICM). These structures typically extend over hundreds of kpc to Mpc and are thought to result from the re-acceleration of fossil relativistic electrons throughout cluster-wide mechanisms, such as merger-induced turbulence. \cite{cuciti22} reported the discovery of more extended radio halos, termed `megahalos', with scales of $\sim 2-3$~Mpc in four galaxy clusters observed with the LOw Frequency ARray (LOFAR) at 53~MHz and 144~MHz, suggesting the presence of relativistic electrons and magnetic fields even in cluster outskirts.
Radio mini-halos are typically observed in the centres of cool-core clusters, with a smaller scale of several hundred kpc. Their origin is still under debate, but one likely scenario involves the re-acceleration of relativistic electrons, injected by central AGN activity, through turbulence caused by gas sloshing in the cluster core \citep{mazzotta08,zuhone13,zuhone15,bravi16,richard20}.  
Recently, the coexistence of a radio mini-halo and an outer giant radio halo has been observed in some cool-core clusters or clusters without significant merger activity \citep{bonafede14,venturi17,savini18,biava21,lusetti24,vanweeren24}, suggesting that past minor or offset mergers could be responsible for powering the re-acceleration of relativistic electrons on a larger scale within the cluster. 
Radio relics are elongated, diffuse radio structures mostly found in the peripheries of galaxy clusters, often aligned perpendicular to the direction of the merger. They trace shock fronts generated during cluster mergers. However, many studies, such as \cite{vanweeren16} and \cite{botteon20a}, have suggested that the standard diffusive shock acceleration (DSA) model \citep[e.g.,][]{BE87}, which accelerates particles from the thermal pool, is insufficient to explain relic luminosities and spectra. 
Additional processes, such as shock re-acceleration, are proposed to play a role in their formation, where merger shocks re-accelerate a pre-existing population of relativistic electrons in the ICM \citep[e.g.,][]{markevitch05,KR11,vanweeren17,botteon20a}. 

In addition to radio halos and relics, large-scale diffuse radio sources known as "radio bridges" have been detected in the outskirts of galaxy clusters, extending along the filamentary structures of the cosmic web. Due to their steep spectra and low surface brightness, radio bridges have only been observed in a limited number of systems, such as between the pre-merger clusters, Abell~399-Abell~401 \citep{govoni19}, Abell~1758 \citep{botteon20c} and a potential candidate of Abell~2061-Abell~2067 \citep{pignataro24}, using LOFAR at $\sim$144~MHz. 
The origin of the inter-cluster radio bridge is not well understood. \cite{govoni19} pointed out that the lifetime of relativistic electrons emitting at 140~MHz is less than 230~Myr, and their maximum travel distance is under 0.1~Mpc \citep{BJ14}, indicating that re-acceleration of in situ relativistic electrons must occur within the radio bridge region. 
Furthermore, \cite{govoni19} suggested that it is difficult to account for the observed strong radio emission from the radio bridges purely assuming acceleration via the observed shock. Therefore, they proposed that there should exist a population of in situ relativistic electrons that have been re-accelerated by shocks and observed under favorable projection effects. 
\cite{BV20} further examined that the turbulence generated during complex dynamic activity (i.e., collapse and accretion) in the bridge can re-accelerate radio plasma seeds injected by past activities (i.e., AGN and star formation), resulting in volume-filling radio emission with steep spectra ($\alpha \sim -1.3 - -1.5$). 
Another type of radio bridge was observed between the clusters and groups, such as the Coma cluster and the NGC 4839 group using LOFAR at 144~MHz \citep{bonafede21} and WSRT at 326~MHz \citep{Kim89,VGF90}. Owing to the highly sensitive MeerKAT L-band data, \cite{venturi22} reported the first GHz detection of a radio bridge between the Abell~3562 cluster and the SC1329-313 group, which was further confirmed by the Australian SKA Pathfinder (ASKAP) at 943.5~MHz \citep{duchesne24}. However, in these two cases, the groups are believed to have experienced off-axis mergers with the cluster and are now at the post-merger status.

Abell~3017 and Abell~3016 (A3017 and A3016 hereafter) are a well-known cluster pair characterized by a prominent X-ray bridge connecting them. The basic properties of two galaxy clusters are presented in Table~\ref{tab:cluster_prop}. This X-ray bridge has been identified by \cite{foex17}, \cite{parekh17}, and \cite{chon19} using Chandra observations. However, the optical density distribution of A3016 was not detected by using the galaxies taken from the SuperCOSMOS and ESO Red catalogues, leading \cite{parekh17} to suggest that A3017 does not form a pair with A3016. This raised the question of whether the X-ray bridge truly connects the two clusters or is only associated with A3017. 
By analyzing VIMOS spectroscopy data and photometric data from the ESO/MPG 2.2m telescope, \cite{foex17} identified four main optical galaxy concentrations: two correspond to clusters A3017 and A3016, one is located between the A3017 and A3016 corresponding to a potential galaxy group, and the another lies $\sim 2.2$~Mpc northeast of A3017, which is being accreted into A3017. The overall optical structure shows an extension from southwest to northeast, spanning at least 4~Mpc, suggesting that the system resides within a large-scale filament. Given the almost zero of the rest-frame velocities of A3016 and the potential group between two clusters, A3016 and the potential group are believed to be bound with A3017, and their alignment is almost horizontal to the plane of the sky. A3017 has likely already experienced a merger with a smaller sub-cluster and is expected to significantly increase in mass through future mergers with surrounding structures. 
\cite{chon19} also confirmed that A3016 is indeed connected with A3017 based on the distribution of the red-sequence galaxies. Moreover, they performed a detailed X-ray analysis and proposed two possible origins for the X-ray bridge between A3016 and A3017. The first scenario suggests that A3016-A3017 is in the post-merger phase, with gas from both clusters contributing to the X-ray bridge as they passed each other. The second possibility is that the X-ray bridge is primarily composed of gas from an embedded galaxy group, and the clusters are still in a pre-merger stage. The temperature of the bridge gas is higher than that of a typical galaxy group, implying additional heating from shocks generated by the approaching clusters and a potential interaction between the A3017 and the embedded group. \cite{chon19} favour the second scenario, as they argue the amount of gas in the bridge is likely too large to be explained solely by the interaction between A3016 and A3017. 
At the A3017 centre, \cite{parekh17} and \cite{pandge21} reported a central AGN with two lobes aligning with the X-ray cavities. The AGN feedback is sufficient to balance the radiative cooling in the cool core of A3017 \citep{pandge21}. Utilizing uGMRT data, \cite{pandge21} reported a radio phoenix at $\sim 150$~kpc of the A3017 cluster centre, with a radio spectral index of $\leq -1.8$, indicating old AGN plasma has been revived due to the shock generated by the cluster merger.

\begin{table}
 \caption{Basic properties of Abell~3016 and Abell~3017.}
 \label{tab:cluster_prop}
 \centering
 \footnotesize
 \setlength{\tabcolsep}{3pt}
 \renewcommand{\arraystretch}{1.2}
 \begin{threeparttable}
 \begin{tabular}{l c c  c c c}
  \hline
       &  R.A. &  Dec.  &  redshift\tnote{a}  & $\rm R_{500}$\tnote{a}   & $\rm M_{500}$\tnote{a}  \\
    &  (J2000) & (J2000) &   &   (Mpc) &   ($\rm 10^{14}~M_{\sun}$)  \\
  \hline
  A3016 &  02h25m22.1s   &  -42$\degree$00$\arcmin$30$\arcsec$  & \multirow{2}{*}{0.2195} &  0.9  &  $1.3_{-0.5}^{+0.9}$  \\
  A3017 &  02h25m52.2s   &  -41$\degree$54$\arcmin$31$\arcsec$  &  &  1.2  &  $3.9_{-1.2}^{+1.9}$  \\
  \hline
 \end{tabular}
 \begin{tablenotes}
    \item [a] The redshift, virial radius, and total mass values are taken from \cite{chon19}. The total masses were derived from measurements of gas surface brightness and temperature with the assumption of hydrostatic equilibrium of the ICM.
 \end{tablenotes}
 \end{threeparttable}
\end{table}

Recently, the MeerKAT Galaxy Cluster Legacy Survey (MGCLS; \citealt{knowles22}) observed the cluster A3017 at 1.28~GHz and detected a radio halo in the cluster centre. Due to A3017's complicated dynamic state and the large-scale filament surrounding it, we re-analyzed the MeerKAT data and combined the Chandra and XMM-Newton archival data to investigate the system further. The paper is organized as follows: in Section~\ref{sect:data}, we describe the procedure of the MeerKAT, MWA, Chandra, and XMM-Newton data reduction; in Section~\ref{sect:result}, we present the imaging and spectral results of the cluster pair A3016-A3017 in both radio and X-ray bands; in Section~\ref{sect:discussion}, we discuss the possible origins of the radio bridge detected between the cluster pair, diffuse radio emission in A3017 and the tadpole-like radio structure; and a conclusion is provided in Section~\ref{sect:conclusion}.
Throughout this paper, we assume the standard flat $\Lambda$CDM cosmology with parameters $H_0=70$~$\rm km~s^{-1}~Mpc^{-1}$ and $\Omega_{m} = 1 - \Omega_{\Lambda}$ = 0.3. At the redshift of A3017 ($z=0.2195$), these values result in a scale of $\sim 213$~$\rm kpc~arcmin^{-1}$ and a luminosity distance of 1093~Mpc. 
For the radio spectral index $\alpha$, we use the convention $S_{\nu} \propto \nu^{\alpha}$, where $S_{\nu}$ is the flux density at the frequency of $\nu$.

\begin{figure*}
    \centering
    \includegraphics[width=0.95\textwidth]{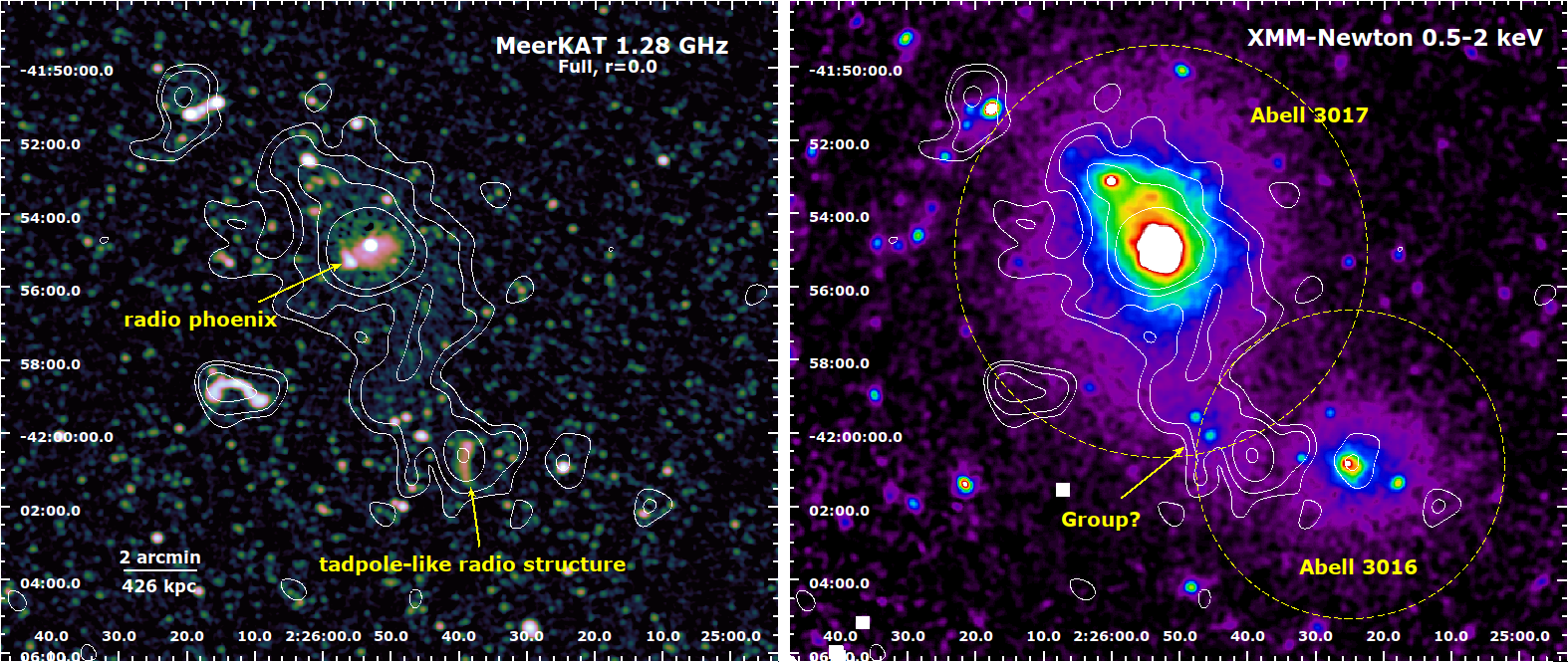}
    \caption{ Left: Full-resolution MeerKAT 1280~MHz image of A3016-A3017 overlaid with the contours of the diffuse radio emission from the low-resolution compact-source-subtracted image (see the top-left panel in Figure~\ref{fig:radio-subbands}). Right: Exposure-corrected XMM-Newton image in the $0.5-2.0$~keV band. The virial radii of A3017 and A3016 are marked. }
    \label{fig:radio-xray}
\end{figure*}

\begin{figure*}
    \centering
    \includegraphics[width=0.45\textwidth]{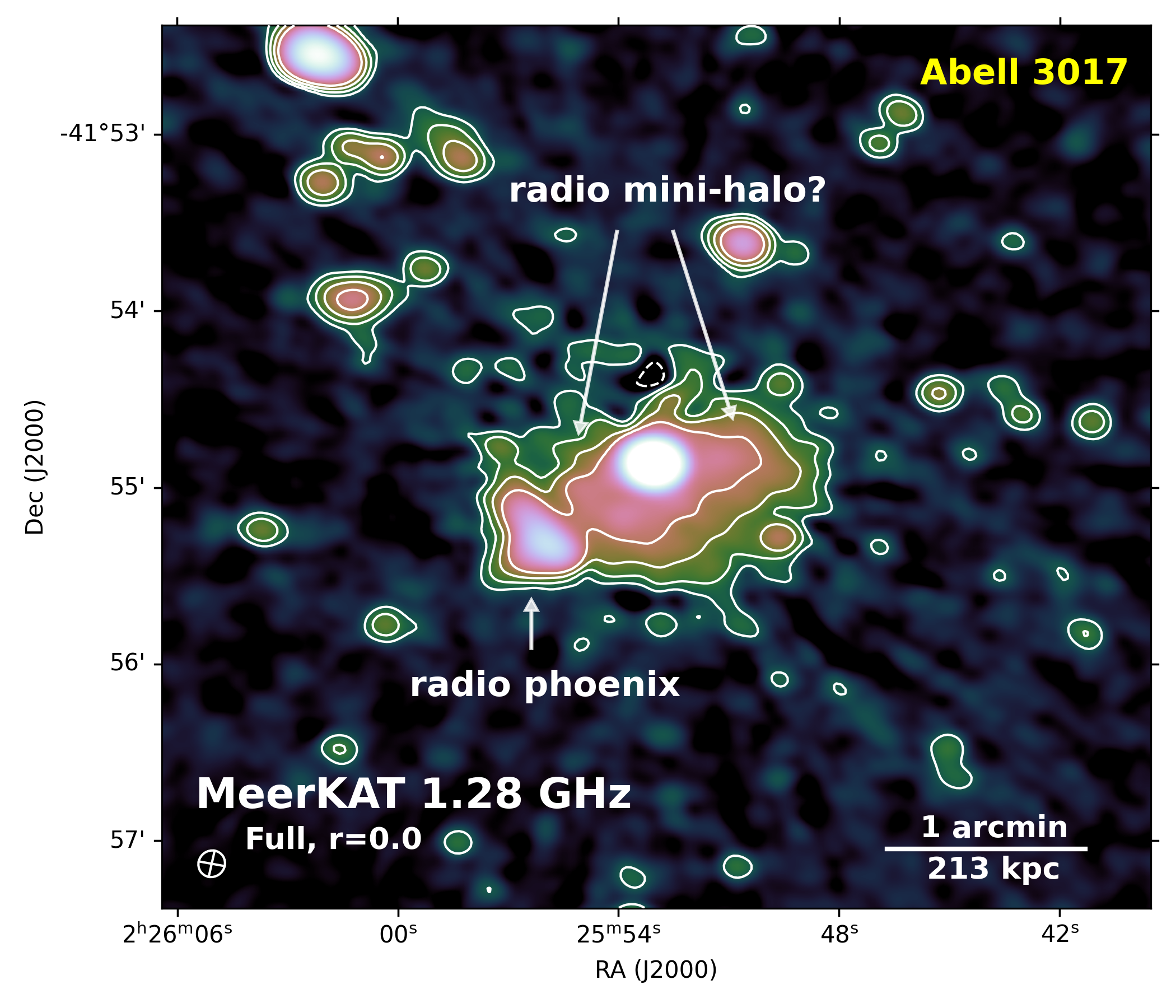}
    \includegraphics[width=0.45\textwidth]{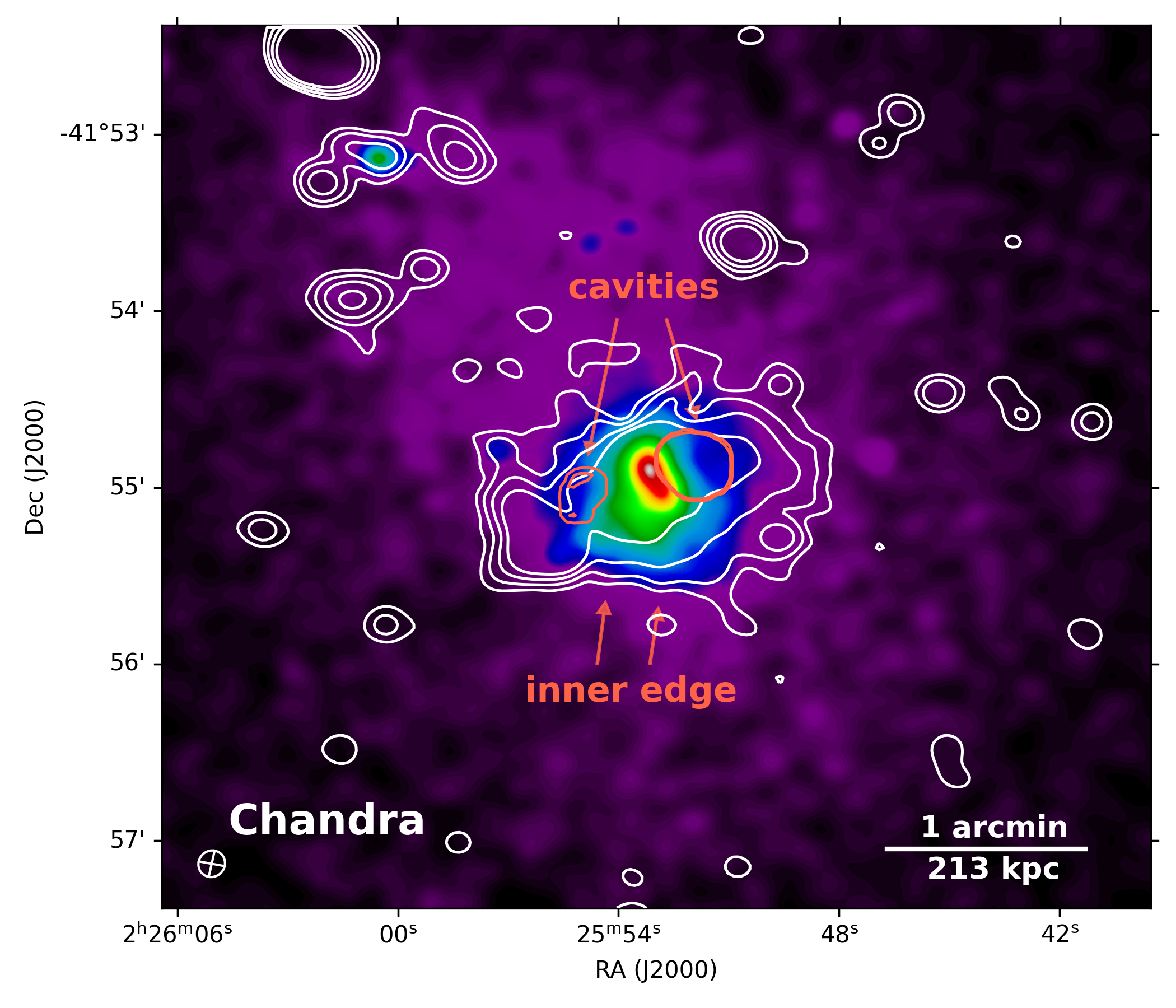}
    \includegraphics[width=0.45\textwidth]{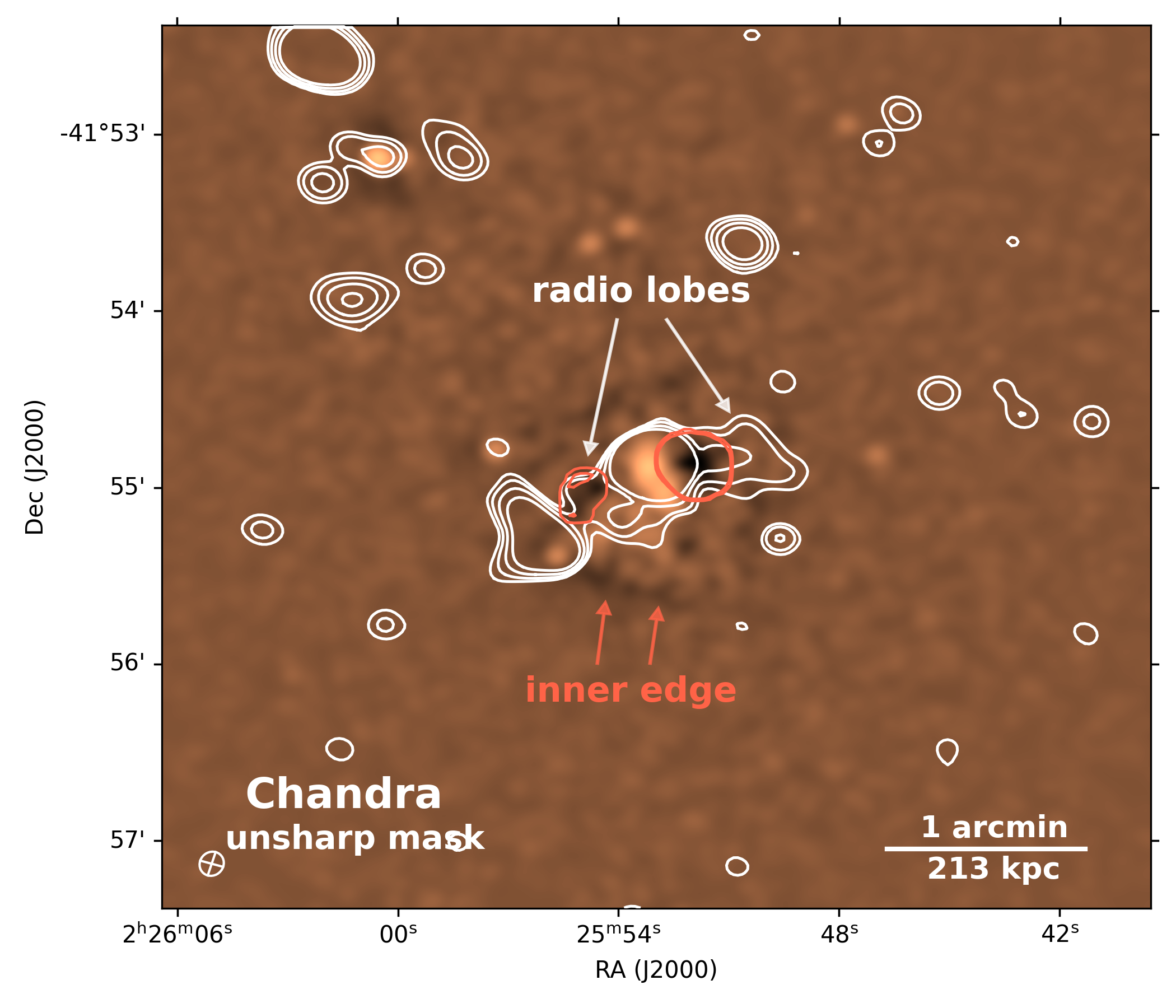}
    \includegraphics[width=0.45\textwidth]{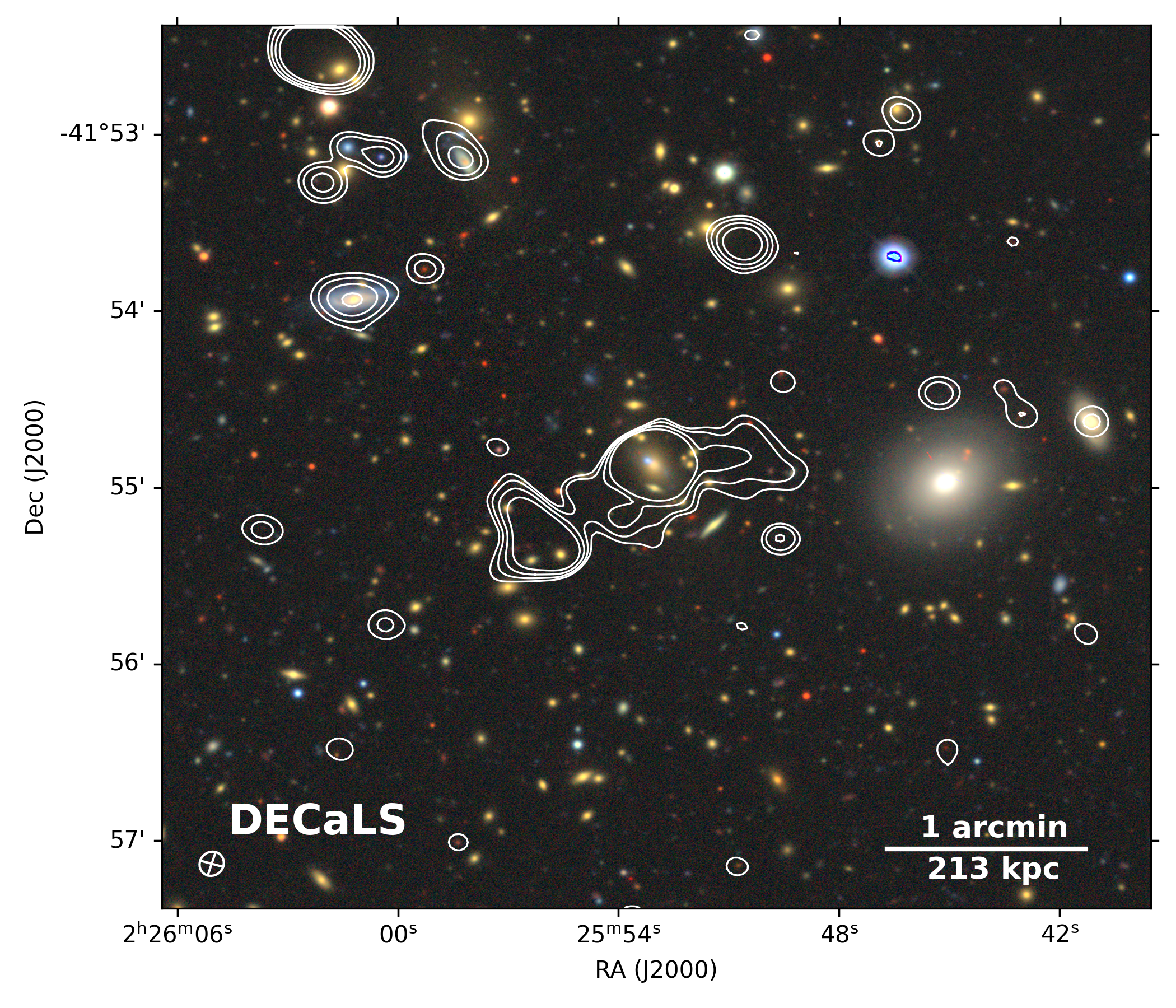}
    \caption{Top-left: Zoomed-in full-resolution MeerKAT 1280~MHz image of the centre region of A3017, with contour levels of [1, 2, 4, 8]$\times 5\sigma$. Negative $-3\sigma$ contours are shown with dotted lines.
    Top-right: Chandra $0.5-7.0$~keV exposure-corrected image of A3017, overlaid with the same contours as in the top-left panel. The red contours represent a pair of X-ray cavities outlined by utilizing the CAvity DEtection Tool (CADET; \citealt{plsek24}). An inner X-ray surface brightness edge induced by gas sloshing is also marked.
    Bottom-left: Chandra $0.5-7.0$~keV unsharp-masked image of A3017, created by subtracting an image smoothed with a 2D Gaussian of 20 pixels from an image smoothed with a 2D Gaussian of 5 pixels. White contours refer to the MeerKAT image of the discrete sources obtained by setting an inner uv-cut of 3437$\lambda$ (corresponding to $\sim 1\arcmin$ or 213~kpc) and show 5$\sigma$, 10$\sigma$, 20$\sigma$, and 40$\sigma$ intensity levels ($\sigma = 7.55 \times 10^{-6}~\rm Jy~beam^{-1}$). 
    Right: Composite DECalS optical image using g-, r-, and z-band data, overlaid with the same contours as in the bottom-left panel.}
    \label{fig:a3017core}
\end{figure*}

\section{Data reduction}
\label{sect:data} 

\subsection{MeerKAT data}
\label{sect:meerkat} 

The cluster A3017 was observed by MeerKAT L-band ($900-1670$~MHz) on 7 January 2019 as part of the MeerKAT Galaxy Cluster Legacy Survey \citep{knowles22}. The total observation time is 10 hours, including the observations of the phase and flux calibrators, i.e., J0203-4349 and J0408-6545.

The initial calibration and self-calibration are performed using \texttt{CARACal} \citep{jozsa20}. Besides the \texttt{AOFlagger} \citep{offringa10,offringa12} and \texttt{Tricolour} \citep{hugo22} introduced in \texttt{CARACal}, we also applied the Common Astronomy Software Applications (\textsc{casa}; \citealt{mcmullin07}) automatic flagging algorithm \texttt{rflag}, to identify and remove the remaining radio frequency interference (RFI). The self-calibration of the target was carried out by two rounds of phase-only self-calibration and two rounds of amplitude and phase self-calibration, with solution intervals of 30s, 15s, 15s, and 15s, respectively. 
We used \texttt{WSclean} \citep{offringa14,offringa17} to generate the clean images using the multi-frequency and multi-scale cleaning methods. Briggs robust 0.0 weighting \citep{briggs95} was used to balance the sensitivity and spatial resolution. All images have been corrected for the MeerKAT L-band primary beam attenuation using the model described in \cite{mauch20}. The final MeerKAT image at 1.28~GHz has a synthesized beam size of $9.08\arcsec \times 8.03\arcsec$ at P.A.$=78.56\degree$, and the root mean squared (rms) noise $\sigma$ of 7.87~$\rm \mu Jy~beam^{-1}$ (see in Table~\ref{tab:propoty}).

\begin{table}
 \caption{Properties of MeerKAT full-band and sub-bands data.}
 \label{tab:propoty}
 \centering
 \footnotesize
 \renewcommand{\arraystretch}{1.2}
 \begin{tabular}{>{\raggedright\arraybackslash}p{0.12\linewidth}>{\centering\arraybackslash}p{0.14\linewidth}>{\centering\arraybackslash}p{0.08\linewidth}>{\centering\arraybackslash}p{0.23\linewidth}>{\centering\arraybackslash}p{0.16\linewidth}}
  \hline
   Image   &  Frequency &  taper & PSF  & $\sigma_{\rm rms}$  \\
    &  (GHz) & ($\arcsec$) &  ($\arcsec \times \arcsec,~\degree$)  &   ($\rm \mu Jy~beam^{-1}$)  \\
  \hline
  Full-res &  1.28  &  --   &  $ 9.08 \times 8.03 $, 78.56  &  7.87 \\
  \hline 
  Discrete &  1.28  &  --   &  $ 8.29 \times 7.26 $, 73.14  &  7.55 \\
  \hline
  Diffuse  &  1.28  &  10   &  $15 \times 15$, 0   &  6.40  \\
           &  1.28  &  40   &  $43 \times 43$, 0   &  29.25  \\
  \hline 
  Diffuse  &  1.01  &  10   &  $15 \times 15$, 0   &   9.35    \\
           &  1.01  &  40   &  $43 \times 43$, 0   &   36.55   \\
  \hline 
  Diffuse  &  1.49  &  10   &  $15 \times 15$, 0   &   6.27     \\
           &  1.49  &  40   &  $43 \times 43$, 0   &   24.25    \\
  \hline
 \end{tabular}
\end{table}

\subsection{MWA data}
\label{sect:mwa}

The A3016-A3017 cluster pair was observed as part of project G0073 using the extended configuration of the Phase II Murchison Widefield Array (MWA-2; \citealt{tingay13,wayth18}). The observing bandwidth was set following the standard GaLactic and Extragalactic All-sky MWA (GLEAM; \citealt{wayth15}) frequency bands: 72$-$103~MHz, 103$-$134~MHz, 139$-$170~MHz, 170$-$200~MHz, and 200$-$231~MHz. 
The data reduction followed the approach described in \cite{duchesne20} and \cite{hurley22}. Since the observation used a 2-min snapshot strategy, each snapshot was processed independently. Briefly, the procedure included: (1) initial calibration was carried out using the in-field sky model -- GLEAM Global Sky Model (GGSM \footnote{\url{https://github.com/nhurleywalker/GLEAM-X-pipeline/tree/master/models}}); (2) Then we imaged each snapshot using \texttt{WSclean} to perform the amplitude and phase self-calibration; (3) Before the deeper cleaning, additional flagging was implemented using the \textsc{casa} with tools of \texttt{tfcrop} and \texttt{rflag} to flag the obvious RFI and unusual channels for each antenna; 
(4) Each snapshot was imaged using \texttt{WSclean} with a Briggs robust $+1.0$ for optimal sensitivity and resolution; 
(5) The astrometric and flux scale corrections were also considered using scripts of \texttt{fits\_warp.py}\footnote{\url{https://github.com/nhurleywalker/fits\_warp}} \citep{HH18} 
and \texttt{flux\_warp}\footnote{\url{https://gitlab.com/Sunmish/flux\_warp}} \citep{duchesne20} for each snapshot; 
(6) After visual inspection, all high-quality snapshots were stacked to produce the final integrated image.  
The properties of MWA-2 images at five frequency bands are summarized in Table~\ref{tab:mwa}. Five MWA-2 images with Briggs robust $+1.0$ weighting are presented in Figure~\ref{fig:mwa5} in Appendix.

\begin{table}
    \caption{Properties of MWA-2 images at five frequency bands.}
    \label{tab:mwa}
    \centering
    \footnotesize
    \renewcommand{\arraystretch}{1.2}
    \begin{tabular}{ccccc}
    \hline
    $\rm Band$ & $\nu_{\rm c}$ & $\rm t_{tot}$ & $\rm PSF$ &  $\sigma_{\rm rms}$  \\
    (MHz) & (MHz) & (hr)  & ($\arcsec \times \arcsec,~\degree$)  &  $(\rm mJy~beam^{-1})$ \\
    \hline
    $\rm {72-103}$   &  $88$   & $18$  & $266.4 \times 183.2$, $-30.0$ & $4.17$ \\
    $\rm 103-134$  &  $118$  & $18$  & $209.6 \times 147.0$, $-29.4$ & $2.04$ \\
    $139-170$  &  $154$  & $18$  & $149.8 \times 103.8$, $-29.2$ & $1.18$ \\
    $170-200$  &  $185$  & $18$  & $121.5 \times 84.8$, $-30.1$ & $0.96$ \\
    $200-231$  &  $216$  & $14$  & $118.9 \times 87.2$, $-27.6$ & $0.86$ \\
    \hline
    \end{tabular}
\end{table}

\subsection{XMM-Newton data}
\label{sect:xmm} 

XMM-Newton data were obtained with the European Photon Imaging Camera (EPIC) on board the XMM-Newton observatory during two pointing observations, i.e., on December 9, 2012 (ObsID: 0692933401; 15.6~ks) and on June 26, 2017 (ObsID: 0803550101; 80.9~ks). 
Data reduction was carried out using the Science Analysis System v20.0.0 (SAS; \citealt{gabriel04}) with the XMM-Newton Extended Source Analysis Software (XMM-ESAS; \citealt{snowden04}). Standard routines \texttt{emchain}, \texttt{mos-filter}, \texttt{epchain} and \texttt{pn-filter} were used to create clean event lists for MOS and pn, respectively. These routines remove periods of anomalously high count rate by calculating Good Time Intervals (GTI) through the \texttt{espfilt} routine. Spectra of all regions were extracted from each XMM-Newton observation using the standard ESAS tools \texttt{mos-spectra}, \texttt{mos\_back}, \texttt{pn-spectra} and \texttt{pn\_back}. The residual soft-proton contamination is explored with the $F_{in}/F_{out}$ ratio, indicating the soft proton contamination ratio between the inside and outside the Field of View (FOV) of the CCDs. MOS2 and pn are slightly contaminated by soft protons in 0692933401, MOS1 is slightly contaminated in 0803550101, and the remaining instruments are otherwise not contaminated.

For each spectrum, we model the full instrumental background components derived from the Filter Wheel Closed (FWC) data as described in \citet{breuer24}. Simply described, the instrumental background consists of continuum emission originating from cosmic rays, described by a broken power-law, and a series of X-ray fluorescence lines originating from the spacecraft itself, described as a series of Gaussian lines at various energies.
The Cosmic X-ray background (CXB) is additionally modelled, described as a combination of foreground and background sources, namely the Local Hot Bubble (LHB), Galactic Halo (GH), a supervirialized component of the Galactic Halo (SV), and unresolved point sources from distant AGN. ROSAT data is used to constrain the LHB temperature, while the other CXB components are determined by creating a blank sky field using the entire FOV of the available data from both Chandra and XMM, excluding the edges of CCDs.
A detailed description of the data processing and background modelling was provided in \cite{hu19} and \cite{breuer24}.

\subsection{Chandra data}
\label{sect:chandra}

Chandra data used in this work were obtained using the Advanced CCD Imaging Spectrometer (ACIS)-I in VFAINT mode on May 1, 2013 (ObsID: 15110; 15~ks) and on April 21, 2015 (ObsID: 17476; 14~ks).
Following the standard procedure suggested by the Chandra X-ray Center, the data was reduced by using the Chandra Interactive Analysis of Observations (\textsc{ciao}; \citealt{fruscione06}) v4.16.0 and Chandra Calibration Database (\textsc{caldb}) v4.10.2. 
The spectra were extracted from each Chandra observation using the \textsc{ciao} tool \texttt{specextract}.
The background construction followed a similar approach to that used for the XMM-Newton. We used the Chandra stowed background, which corresponds to the XMM-Newton FWC background, and modelled its continuum emission and the instrumental lines with a similar combination of a power law and a series of Gaussian lines. Detailed data reduction and analysis descriptions are described in \cite{hu19} and \cite{breuer24}.

In this work, all spectra from both Chandra and XMM-Newton were optimally binned using the method described in \citet{optbin}, and all astrophysical parameters were linked through all spectra for a joint fit using \textsc{xspec} v12.13.0 and \textsc{atomdb} v3.0.9.

\section{Results}
\label{sect:result} 

\subsection{Radio morphology}
\label{sect:radio_image}

\begin{figure*}
    \centering
    \includegraphics[width=0.33\textwidth]{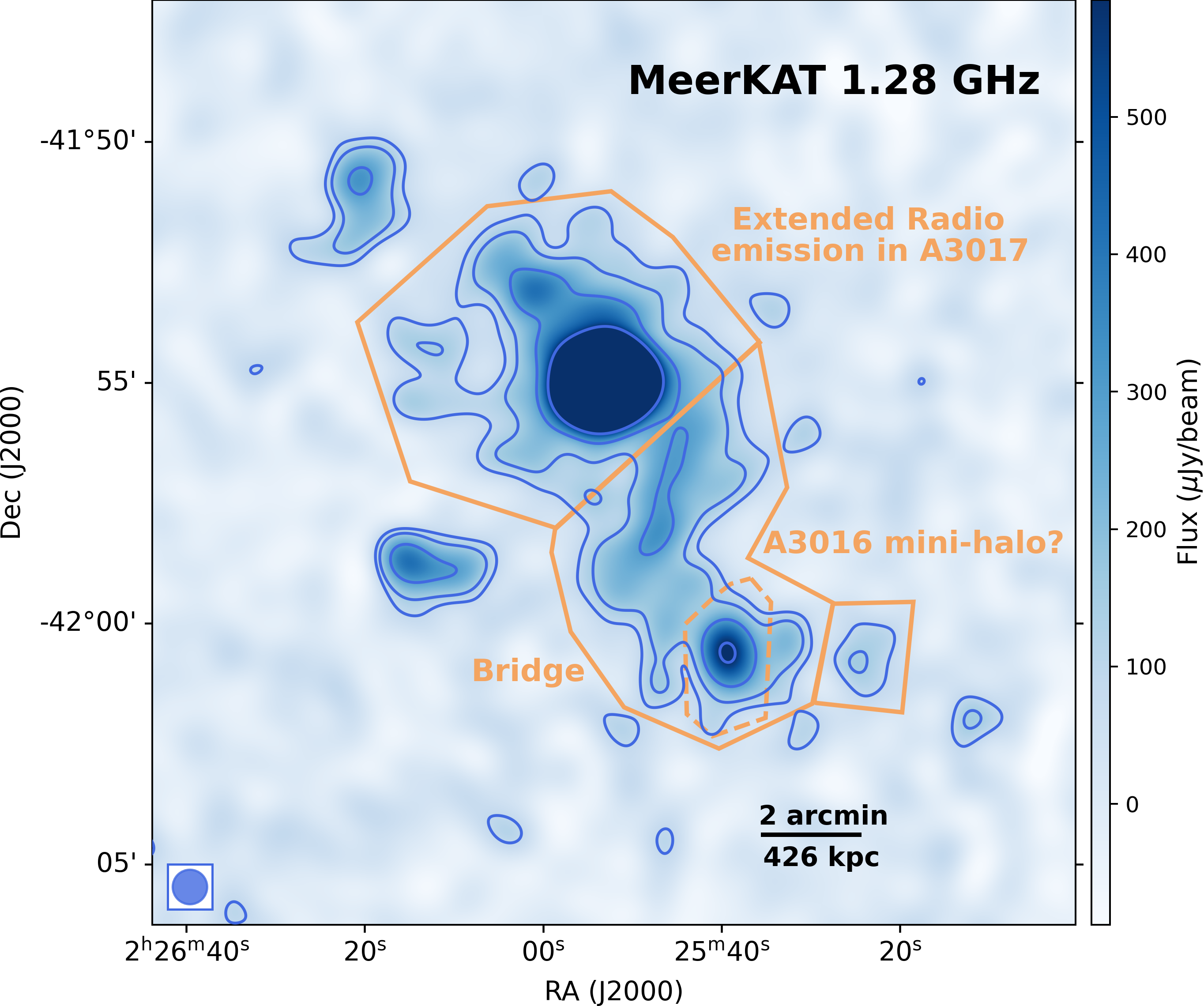}
    \includegraphics[width=0.33\textwidth]{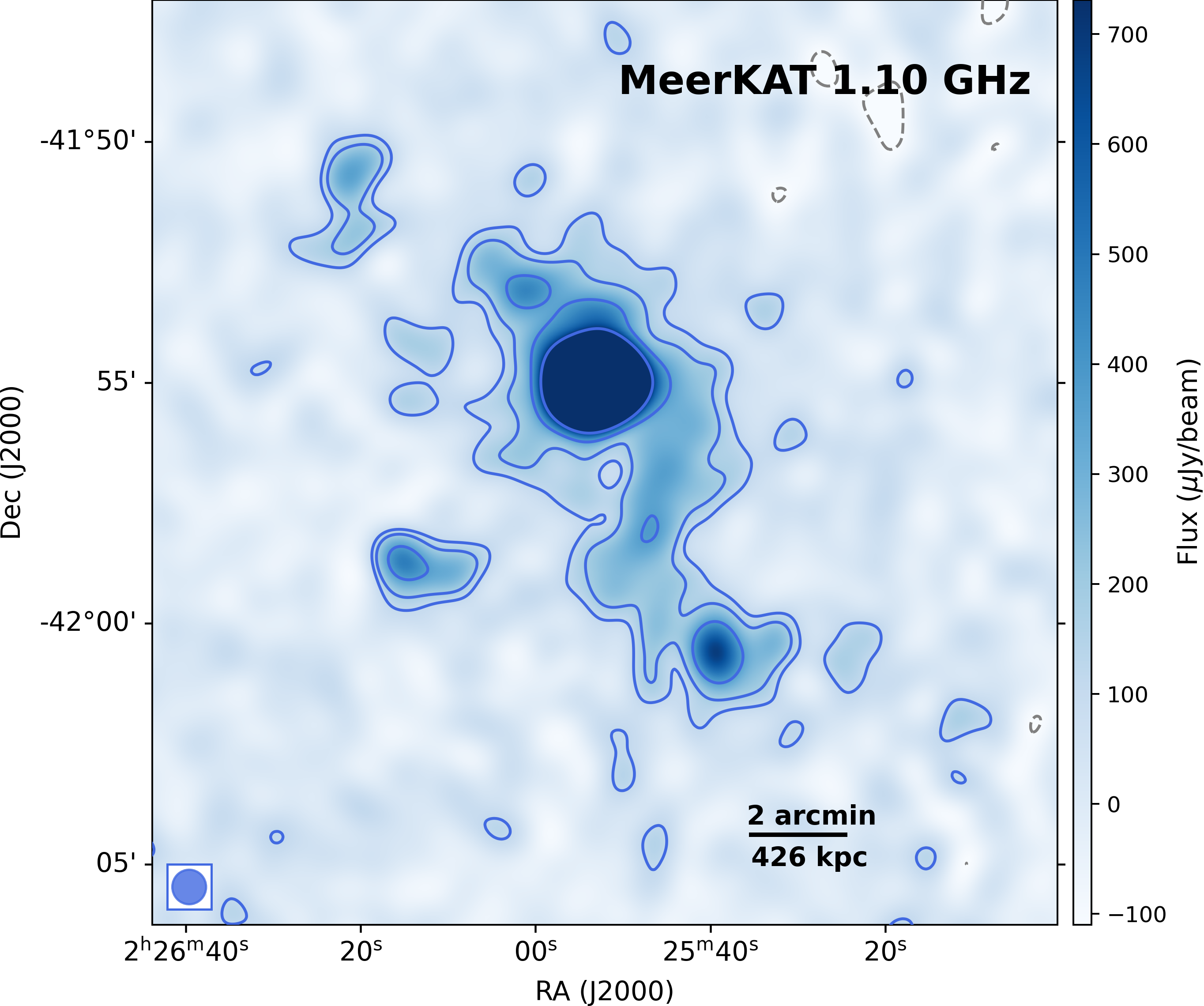}
    \includegraphics[width=0.33\textwidth]{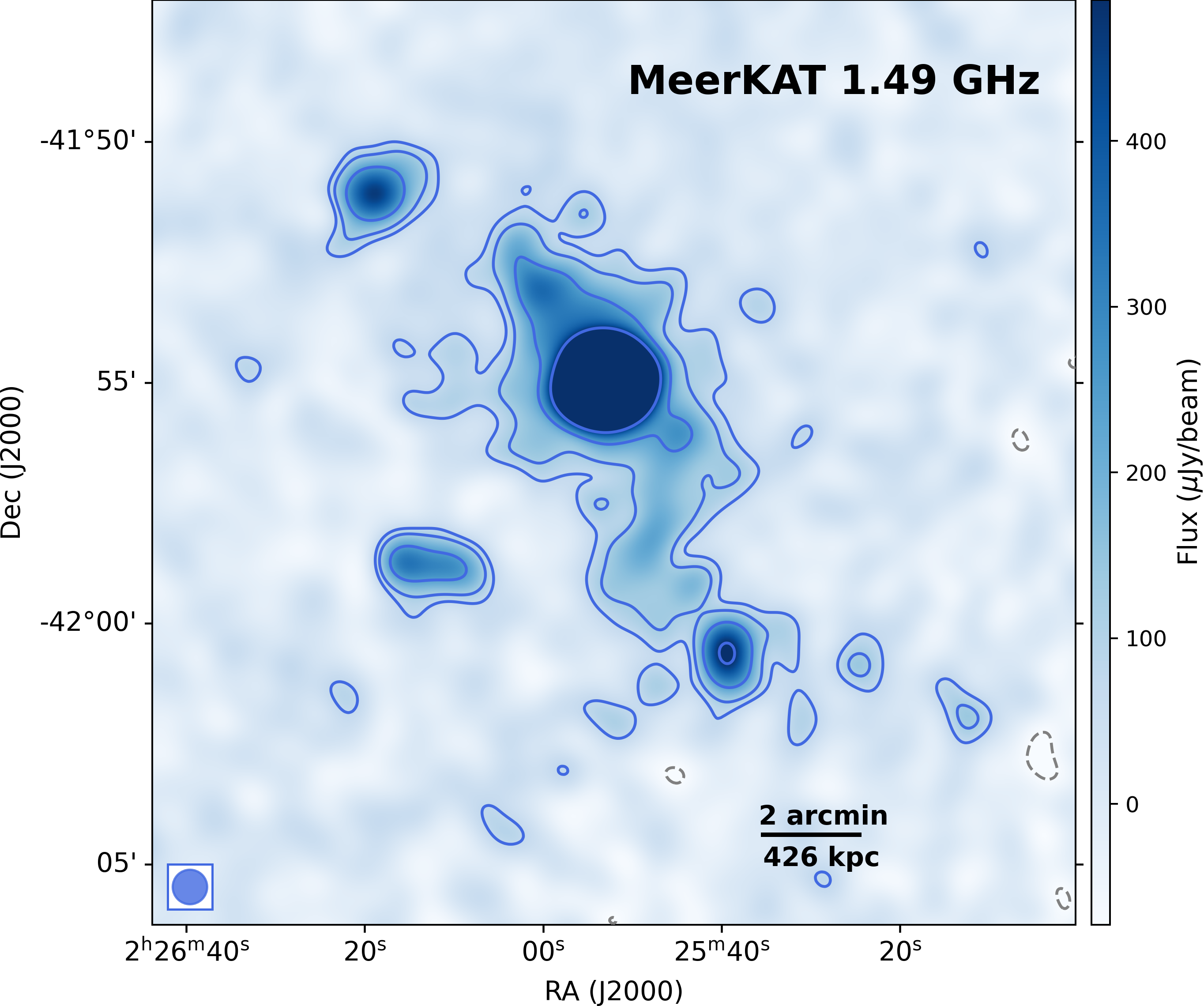}
    \includegraphics[width=0.33\textwidth]{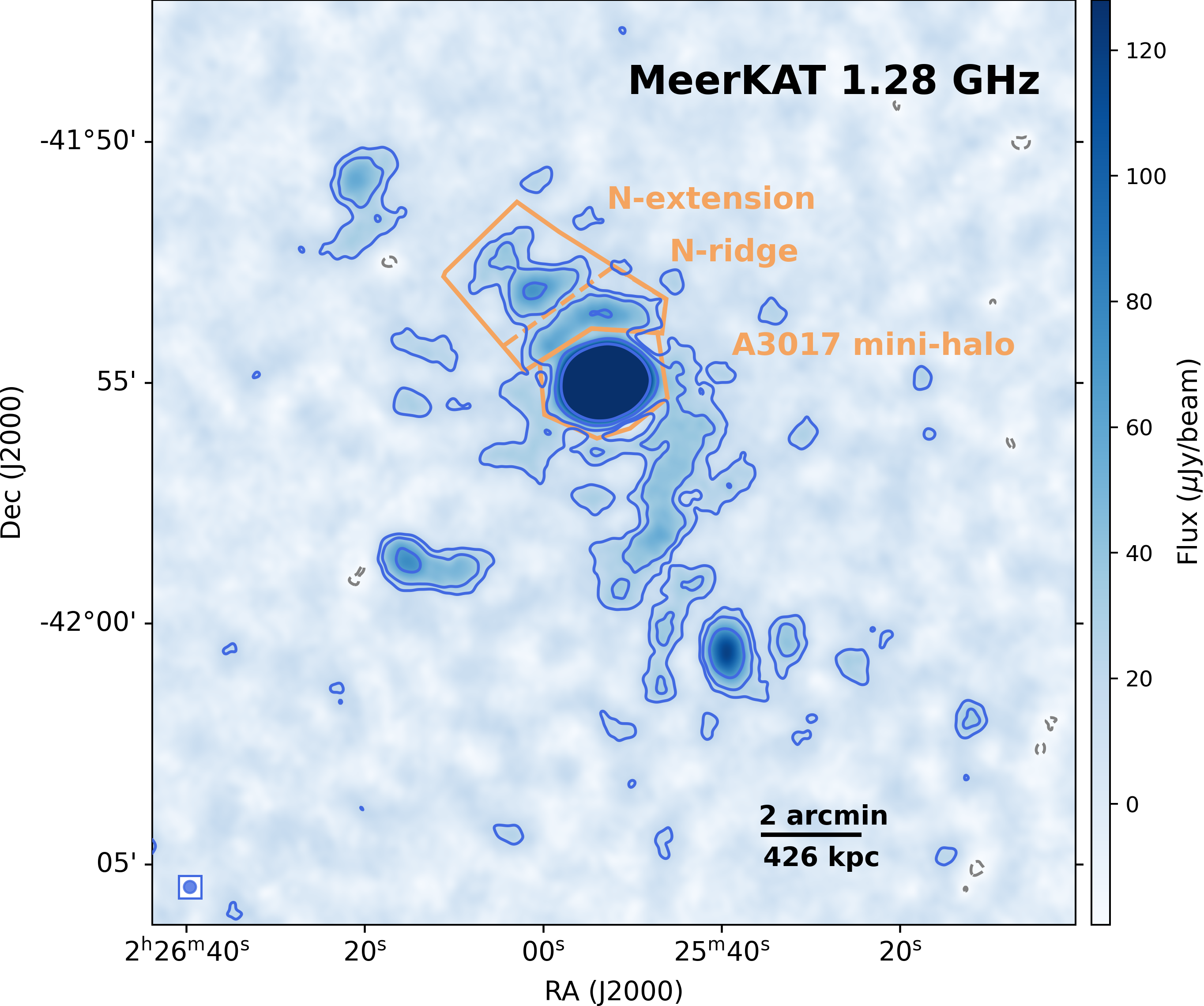}
    \includegraphics[width=0.33\textwidth]{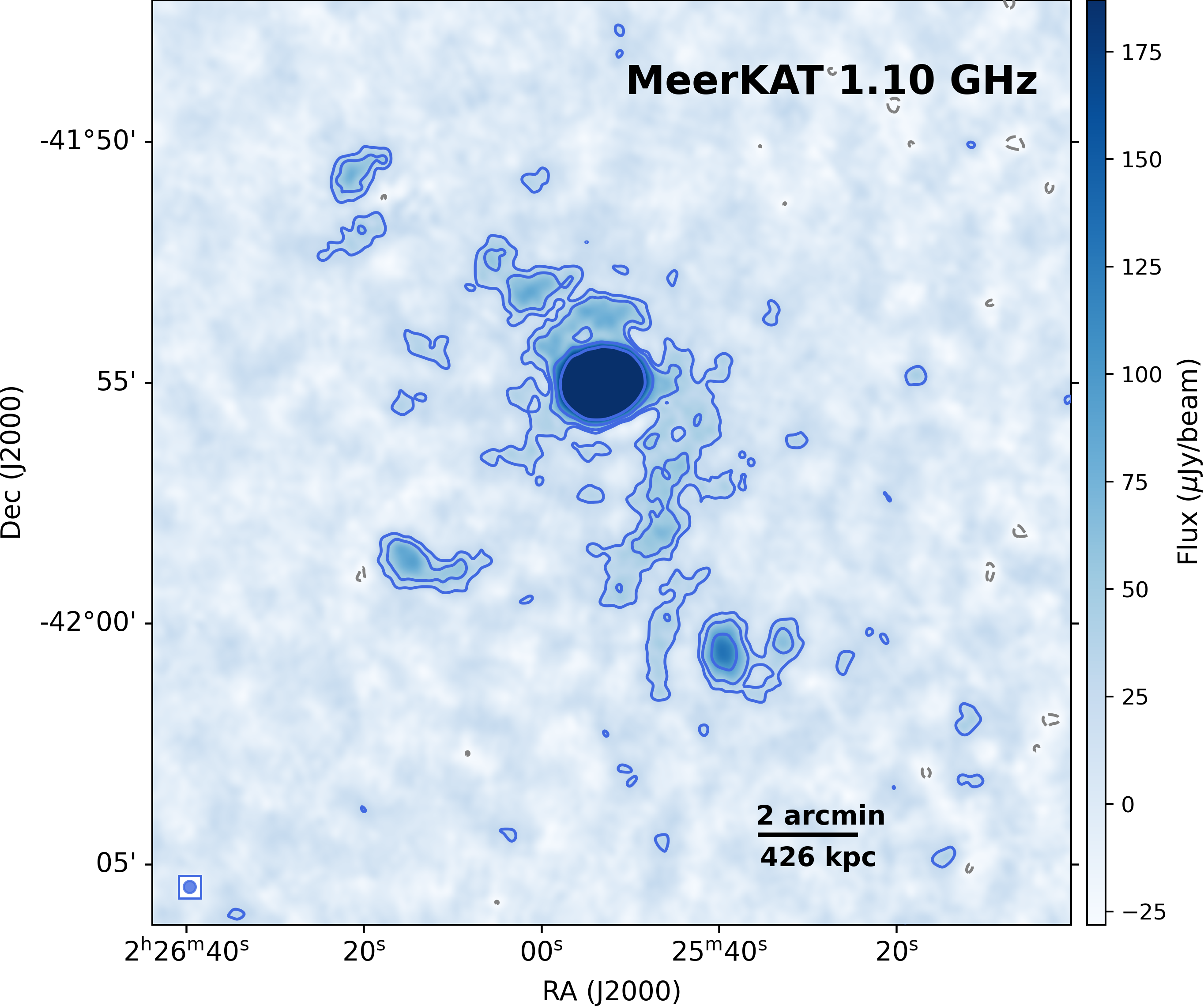}
    \includegraphics[width=0.33\textwidth]{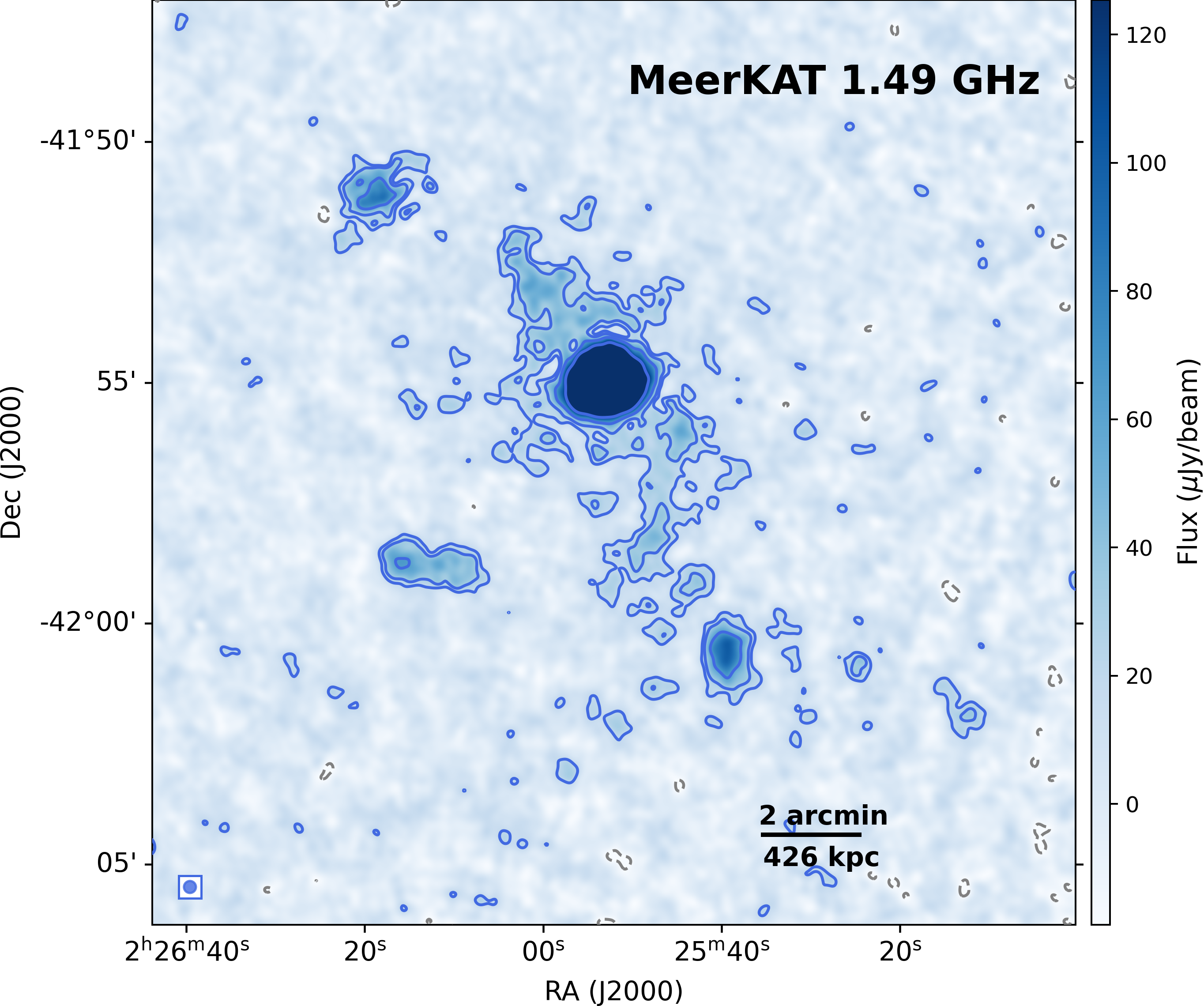}
    \caption{Top: Low-resolution ($43\arcsec \times 43\arcsec$) compact-source-subtructed MeerKAT images of diffuse radio emission (unit: $\rm Jy~beam^{-1}$) at full-band and two subbands with central frequencies of 1.28~GHz, 1.10~GHz and 1.49~GHz, respectively. Bottom: Same as the top, but at a higher resolution ($15\arcsec \times 15\arcsec$). Contours show 3$\sigma$, 5$\sigma$, 10$\sigma$, and 20$\sigma$ intensity levels. Negative $-3\sigma$ contours are shown with grey dashed lines. The regions used for estimating the flux densities of each substructure are indicated.  }
    \label{fig:radio-subbands}
\end{figure*}

\begin{figure*}
    \centering
    \includegraphics[width=0.32\textwidth]{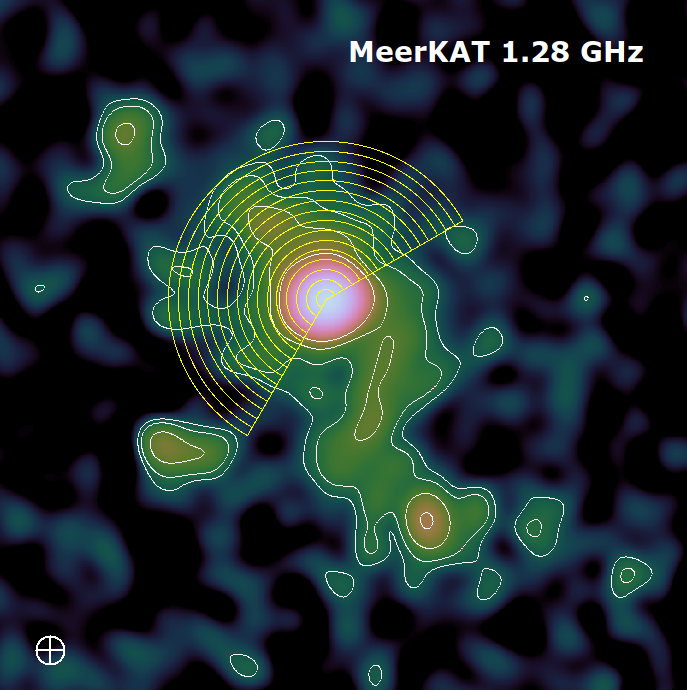}
    \hspace{0.2cm}
    \includegraphics[width=0.48\textwidth]{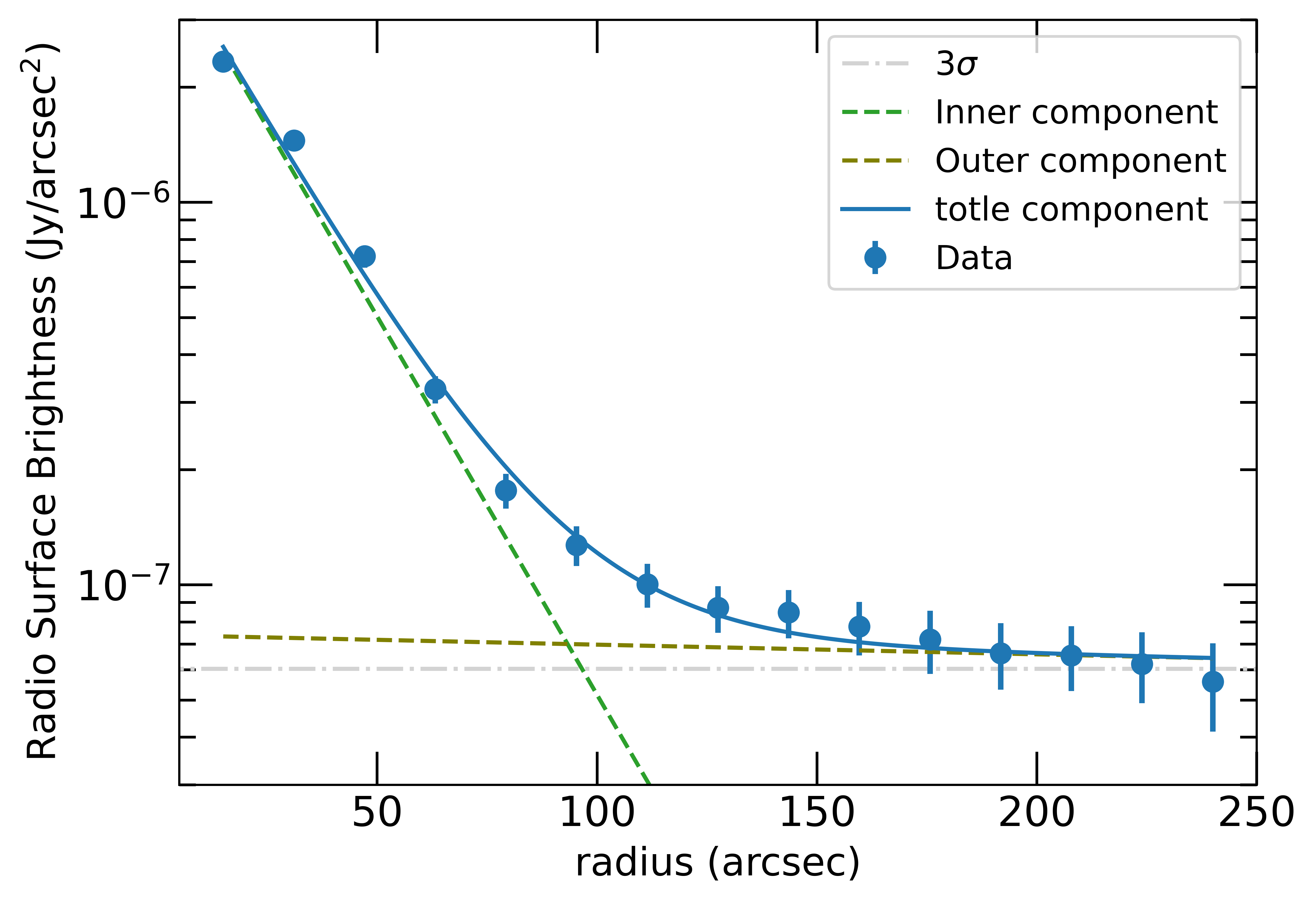}
    \caption{ Left: The region used to extract the surface brightness from the compact-source-subtracted low-resolution MeerKAT 1280~MHz image.
    Right: Radio surface brightness profile, with the solid line representing the best-fit result using a double exponential model. The two individual exponential components are shown as dashed lines, and the horizontal dash-dotted line indicates the $3\sigma$ in the compact-source-subtracted low-resolution MeerKAT image.}
    \label{fig:radio-sbp}
\end{figure*}

\begin{figure}
    \centering
    \includegraphics[scale=0.3]{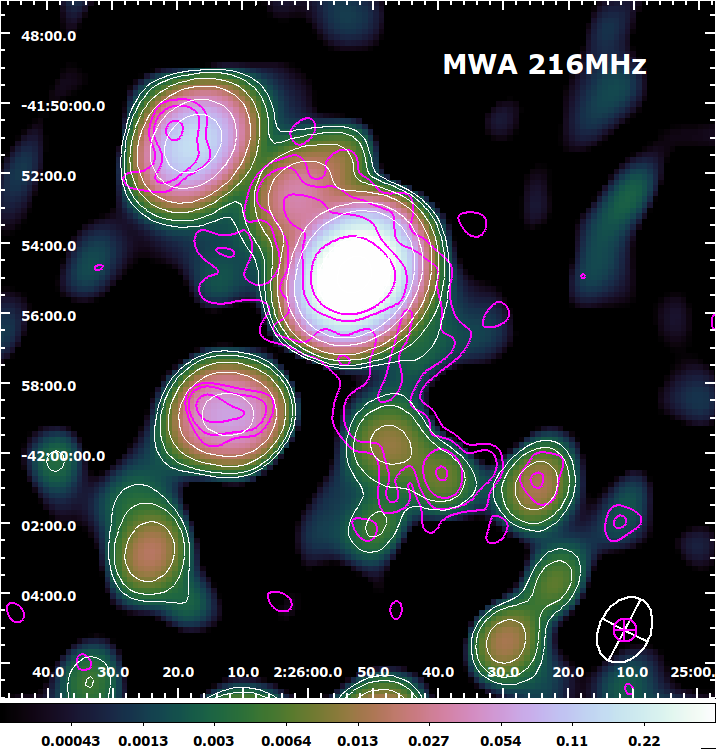}
    \caption{MWA-2 216~MHz image of A3016-A3017 (unit: $\rm Jy~beam^{-1}$).  The overlaid white contours are from MWA-2 216~MHz images with the level of [3, 5, 10, 20, 50, 100]$\times \sigma$. The magenta contours are from the compact-source-subtracted low-resolution MeerKAT 1280~MHz image of diffuse radio emission.  }
    \label{fig:mwa216mhz}
\end{figure}

The full-resolution MeerKAT image of the cluster pair A3016-A3017 at 1280~MHz with robust 0.0 weight is shown in the left panel of Figure~\ref{fig:radio-xray}. 
At the centre of A3017 (see Figure~\ref{fig:a3017core}), an extended radio emission spanning $\sim 1\arcmin.3$ (i.e., $\sim 277$~kpc) surrounds the central AGN, including the two radio lobes previously detected by GMRT at $300-600$~MHz \citep{parekh17,pandge21}. However, the emission appears more extended and is confined within the X-ray bright core, indicating a potential existence of the radio mini-halo. Specifically, we note that this extended radio emission spatially corresponds to, but is bound within the inner X-ray surface brightness edge revealed in both the Chandra exposure-corrected image and unsharp-masked image (see Figure~\ref{fig:a3017core}), further supporting the presence of a radio mini-halo powered by gas sloshing \citep{MG08,zuhone13}. A nearby radio phoenix located $\sim 40\arcsec$ southeast of the central AGN identified by GMRT at $300-600$~MHz \citep{pandge21} is also clearly seen in the top-left panel of Figure~\ref{fig:a3017core}. 

\cite{knowles22} first reported a radio halo detected in A3017 with a largest angular size of $2\arcmin.5$ (i.e., 510 kpc). To further investigate the radio min-halo and radio halo, and uncover potential diffuse radio emission, we followed the standard approach of imaging only diffuse emission by subtracting discrete sources, a method commonly used in many studies using different instruments \citep[e.g.,][]{botteon18,duchesne21,riseley23,sikhosana23,botteon24,duchesne24,riseley24}. The basic procedure is summarized as follows: first, we imaged the discrete sources by setting an inner uv-cut of 3437$\lambda$, corresponding to approximately 1$\arcmin$ (or 213~kpc physical scale).
We carefully compared the compact source models by using different uv-cuts, i.e., 1$\arcmin$, 1$\arcmin$.5, and 2$\arcmin$, and found that the uv-cut of 1$\arcmin$ is a relatively reasonable choice as it basically accounts for the central AGN and radio lobes of A3017 and minimizes the removal of small-scale diffuse emission. The MeerKAT image of the central AGN and radio lobes by setting the inner uv-cut of 1$\arcmin$ is shown as white contours in the bottom panels in Figure~\ref{fig:a3017core}. The radio lobes align well with the X-ray cavities\footnote{In Figure~\ref{fig:a3017core}, the X-ray cavities are clearly visible in the Chandra $0.5-7.0$~keV image and are further outlined using the CAvity DEtection Tool (CADET; \citealt{plsek24}), an automatic machine learning pipeline designed to detect cavities from broad-band Chandra flux images.} revealed on the Chandra $0.5-7.0$~keV images (see the top-right and bottom-left panels in Figure~\ref{fig:a3017core}) and are consistent with the radio lobes detected by GMRT at $300-600$~MHz \citep{parekh17,pandge21}.

Then, after subtracting the models of discrete sources from the visibility, the diffuse radio emission was imaged at different resolutions. 
At the higher resolution of $15\arcsec$, we achieved the rms of about 6.40~$\mu$Jy. To provide better sensitivity to extended low surface brightness features, we also produced the images at lower resolution ($43\arcsec$) by setting a Gaussian taper of $40\arcsec$. The compact-source-subtracted images at two resolutions are exhibited in Figure~\ref{fig:radio-subbands}, and the basic properties of these images are presented in Table~\ref{tab:propoty}. 
For comparison, we also convolved the MeerKAT image (before the compact-sources subtraction) to $15\arcsec$ and $43\arcsec$ resolutions and present them in Figure~\ref{fig:convlved} in Appendix.

In Figure~\ref{fig:radio-subbands}, the diffuse radio emission is clearly visible in MeerKAT 1280~MHz images at two different resolutions. The diffuse radio emission has several prominent substructures:
\begin{enumerate}
    \item Radio bridge: The most striking structure is the radio bridge extending from A3017 toward A3016, spanning $\sim 1.2$~Mpc (i.e., $5\arcmin.6$) in length, in projection. This radio bridge aligns well with the X-ray bridge (see the right panel in Figure~\ref{fig:radio-xray}). A small blob within the radio bridge coincides with the residuals of the tadpole-like radio structure visible in the left panel in Figure~\ref{fig:radio-xray}. This tadpole-like structure was first identified by \cite{parekh17} in GMRT observations at 235~MHz and 610~MHz, and will be further discussed in Section~\ref{sect:radio-vertical}.
    \item Extended radio emission in A3017: The extended radio emission with a scale of $\sim 1$~Mpc (i.e., $\sim 5\arcmin$) at low resolution is centrally located at the A3017 core, and extends toward the northeast, corresponding with the X-ray extension of A3017. The high-resolution compact-source-subtracted image (see the bottom panels of Figure~\ref{fig:radio-subbands}) reveals more distinct substructures, suggesting that the extended radio emission may be composed of a radio mini-halo at the core and a large outer radio halo with a northern extension, within which a relic-like ridge is embedded: 
    \begin{itemize}
        \item Radio mini-halo: The brightest diffuse component shows a relatively elliptical morphology, closely aligned with the bright X-ray core of A3017, indicating the presence of a radio mini-halo. 
        \item Northern radio extension (N-extension): To the north of the mini-halo, a radio extension coincides with the northern X-ray excess of A3017 (Figure~\ref{fig:radio-xray}). The N-extension could represent the brightest part of an outer radio halo, likely related to turbulent gas motion from ongoing merging activity within A3017. 
        \item Northern ridge (N-ridge): Within the N-extension, we notice an arc-shaped radio ridge located north of the A3017 core in the high-resolution image. The relic-like feature has an angular extent of $\sim 2\arcmin.2$ (i.e., $\sim 460$~kpc) and a projected distance of $\sim 1\arcmin.4$ (i.e., $\sim 300$~kpc or $0.25\rm R_{500}$) from the centre of A3017. This relic-like ridge may indicate a central radio relic. Unlike typical radio relics, which are usually found at the cluster outskirts ($>0.2-0.5\rm R_{200}$), central radio relics are much closer to the cluster core, and are rarely detected due to their weak radio emission, likely caused by the lower energy dissipation of shocks in the cluster centres \citep{vazza12}. It is also possible that the relic-like ridge originated from previous AGN activity, but its size is much larger than the typical scales of the radio lobes, and no corresponding X-ray cavity has been detected. 
    \end{itemize}
    \item Radio mini-halo of A3016? The diffuse radio emission associated with A3016 was marginally ($3\sigma$) detected at low resolution, spanning $1\arcmin.1$ (i.e., $\sim 234$~kpc). Since A3016 shows no clear signs of a merger and its brightest cluster galaxy (BCG) hosts a powerful AGN, this emission could be a candidate for a mini-halo. Further observations are required to examine its existence. 
\end{enumerate}

To further examine the radio mini-halo and outer halo of A3017, we extracted the surface brightness from the compact-source-subtracted low-resolution MeerKAT 1280~MHz image along the northeast direction, avoiding contamination from the radio bridge, and collected only pixels with values greater than $3\sigma$. The selected region and radio surface brightness profile are presented in Figure~\ref{fig:radio-sbp}.
The resulting profile shows a steep inner component and a relatively flat outer component. To characterize the two components, we followed the method described in \cite{vanweeren24} and applied a double exponential model \citep{murgia09} to fit the radio surface brightness profile, expressed as: 
\begin{equation}    
I_{\rm R}(r) = I_{\rm 0, inner}e^{-r/r_{\rm e, inner}} + I_{\rm 0, outer}e^{-r/r_{\rm e, outer}} 
\end{equation}
where $I_{\rm 0, inner}$ and $I_{\rm 0, outer}$ are the central surface brightness values of the inner and outer components, and $r_{\rm e, inner}$ and $r_{\rm e, outer}$ are their $e$-folding radii.
The fitting result suggests that the inner component has a central surface brightness $I_{\rm 0, inner} = 4.9 \pm 0.4$~$\rm \mu Jy~arcsec^{-2}$ and $r_{\rm e, inner} = 21\arcsec.9 \pm 1\arcsec.1$ (i.e., $78 \pm 4$~kpc), consistent with values of $r_{\rm e, inner}$ from other radio mini-halos \citep[e.g.,][]{murgia09,lusetti24,vanweeren24}. 
However, the outer component is quite faint and flat, with $I_{\rm 0, outer} = 0.07 \pm 0.03$~$\rm \mu Jy~arcsec^{-2}$ and a large, poorly constrained $r_{\rm e, outer}$. This may be attributed to the limited number of data points as they are approaching the $3\sigma$ threshold, and potential contamination from small-scale substructures, such as the N-ridge. Nevertheless, the presence of the outer component suggests that it may be related to another scenario involving the N-extension, which will be discussed in Section~\ref{sect:radio-halo}.
It is worth noting that the surface brightness of the mini-halo could be influenced by the central AGN and radio lobes. As mentioned in \cite{bonafede23}, subtracting the AGN and radio lobes from the visibility data is a more effective approach for investigating the mini-halo; however, challenges such as under- or over-subtraction cannot be entirely avoided. Further radio observations with higher spatial resolution, across different frequency bands, and accompanied by precise spectral index maps, are essential for more effectively distinguishing the emission from the AGN, radio lobes, and mini-halo.

It is also important to investigate the presence of the radio bridge at low frequencies. In Figure~\ref{fig:mwa216mhz}, the radio emission observed with MWA-2 at 216~MHz generally aligns with the diffuse structures observed in the compact-source-subtracted low-resolution MeerKAT 1280~MHz image. However, the radio bridge between the two clusters is not clearly detected by MWA-2 data. The non-detection is likely due to the limited sensitivity of MWA-2, with an rms level of approximately 0.86~$\rm mJy~beam^{-1}$ at 216~MHz.
Given the relatively large PSF of MWA-2 data and the uncertainties in the spectral indices of discrete sources at MWA frequencies, we were unable to obtain accurate flux density measurements for the diffuse radio structures from the MWA-2 results. Consequently, only the MeerKAT data were used for the spectral analysis.

\subsection{Radio spectral analysis}
\label{sect:radio_spc}

\begin{table*}
 \caption{Flux densities of diffuse radio structures at MeerKAT different frequencies.}
 \label{tab:spectra}
 \centering
 \footnotesize
 \renewcommand{\arraystretch}{1.2}
 \begin{adjustbox}{width=0.9\textwidth, center}
 \begin{threeparttable}
 \begin{tabular}{l |c c | c c | c c}
  \hline
    &  \multicolumn{2}{c|}{Full-band}   &  \multicolumn{2}{c|}{Two subbands}  &   \\
  \hline
    &  $S_{1280~\rm MHz}$  &  $L_{1280~\rm MHz}$ & $S_{1101~\rm MHz}$ & $S_{1490~\rm MHz}$ & $\alpha_{1101~\rm MHz}^{1490~\rm MHz}$ & $\alpha_{\rm fit}$ \\
     & (mJy) &   ($\rm 10^{23}~W~Hz^{-1}$)  &   (mJy)  &  (mJy)  &   &  \\
  \hline
  Bridge\tnote{a}  &  $4.27 \pm 0.49$   &  6.09  &  $5.01 \pm 0.57 $ & $2.62 \pm 0.31$  &  $-2.14 \pm 0.54$  & $-2.13 \pm 0.52$   \\
  \hline 
  Extended radio emission in A3017\tnote{a} & $ 24.00 \pm 1.35 $ &   34.23  & $27.77 \pm 1.56$ & $17.38 \pm 0.98 $ &  $-1.55 \pm 0.26$ &  $-1.55 \pm 0.26$    \\
  \hline   
  Mini-halo\tnote{b}  &  $ 17.09 \pm 0.90 $ & 24.38 &  $ 20.47 \pm 1.09 $ & $ 12.66 \pm 0.68 $ &  $ -1.59 \pm 0.25 $  & $ -1.59 \pm 0.25  $ \\
  \hline
  N-extension\tnote{b}  &  $ 2.69 \pm 0.18 $ & 3.84 &  $ 2.75 \pm 0.19 $ & $ 2.11 \pm 0.14 $ &  $ -0.88 \pm 0.32 $  & $ -0.88 \pm 0.31 $ \\
  \hline
  N-ridge\tnote{b}   & $ 1.34 \pm 0.10 $  & 1.91  &  $ 1.46 \pm 0.11 $ & $ 1.07 \pm 0.09 $ &  $ -1.03 \pm 0.37 $  &  $ -1.01 \pm 0.36 $  \\
  \hline
  A3016 mini-halo\tnote{a} &  $ 0.26 \pm 0.04 $ &  0.37 &  $0.31 \pm 0.05$ & $0.14 \pm 0.02$ & $-2.63 \pm 0.71$  & $-2.67 \pm 0.67$ \\ 
  \hline
 \end{tabular}
 \begin{tablenotes}
      \item [a] The measurements of flux density were derived from the compact-source-subtracted, low-resolution image. The residual of the tadpole-like radio structure was subtracted when measuring the radio bridge.
      \item [b] The measurements of flux density were derived from the compact-source-subtracted, high-resolution MeerKAT image. 
 \end{tablenotes}
 \end{threeparttable}
 \end{adjustbox}
\end{table*}

We selected three polygon regions to estimate the integrated flux densities of the total extended radio emission of A3017, radio bridge and A3016 mini-halo from the compact-source-subtracted low-resolution MeerKAT image. Additionally, three smaller polygons were chosen to account for the smaller-scale substructures, i.e., the mini-halo, N-extension, and N-ridge, revealed in the compact-source-subtracted high-resolution MeerKAT image. The selected regions are presented in Figure~\ref{fig:radio-subbands}. Note that the flux density for the radio bridge was obtained after subtracting the contribution from the residual of the tadpole-like radio structure.
The flux densities of these diffuse substructures were measured from the pixels where the emission is above 3$\sigma$.
The error on the flux density ($\sigma_{S_{\nu}}$) was estimated taking into account the image noise and calibration uncertainty as follows,
\begin{align}
    \sigma_{S_{\nu}} = \sqrt{(N_{\rm beam} \times \sigma_{\rm rms}^{2}) + (f \times S_{\nu})^{2}} 
    \label{eq_err}
\end{align}
where $N_{\rm beam}$ is the number of beams covering the entire region of interest, $\sigma_{\rm rms}$ is the local rms noise of the image, and $f$ is the flux scale uncertainty. Here, we considered $5\%$ uncertainty as suggested in \cite{knowles22}.
The flux density results of diffuse radio structures are presented in Table~\ref{tab:spectra}.

To have a basic understanding of the spectral indices of these diffuse structures, we separated the L-band into two subbands, 890~MHz $-$ 1310~MHz and 1310~MHz $-$ 1670~MHz, and then generated images at the central frequencies of 1101 MHz and 1490 MHz, respectively. Both in-band MeerKAT images were convolved to the same resolution ($15\arcsec \times 15\arcsec$ and $43\arcsec \times 43\arcsec$), and re-grided to have the same pixel scale (see Figure~\ref{fig:radio-subbands}). The measurements of rms ($\sigma$) from these images are summarized in Table~\ref{tab:propoty}. Most of the diffuse radio emission is visible in the subband MeerKAT images, providing an opportunity to estimate the spectral index of the radio bridge for the first time. Using the same polygon regions, we estimated the flux densities of these diffuse substructures from the two subband images and calculated their spectral indices. All results are presented in Table~\ref{tab:spectra}.

The in-band spectral index of the A3017 mini-halo is $\alpha_{\rm fit}^{\rm mini-halo} = -1.59\pm 0.25$, which is steep but falls within the typical range of $-1$ to $-1.5$ observed for mini-halos \citep[e.g.,][]{gitti04,venturi17,giacintucci19}. We note that the steep spectra of the A3017 mini-halo may be influenced by the AGN lobes, and future observations are needed to distinguish the contributions from the AGN lobes and mini-halo. 
The spectral index of the N-extension appears relatively flat, with $\alpha_{\rm fit}^{\rm N-extension} = -0.88\pm 0.31$. 
The N-ridge has a spectral index of $\alpha_{\rm fit}^{\rm N-ridge} = -1.01\pm 0.36$, which is consistent with typical values of radio relics reported in the literature \citep[e.g.,][]{bonafede12,rajpurohit21}.  
For the radio bridge, the spectral index of the entire region is $\alpha_{\rm fit}^{\rm bridge} = -2.13\pm 0.52$. 
Note that the frequency range (1101~MHz$-$1490~MHz) used to determine the spectral characterisation of the diffuse radio emission in this work is relatively narrow. Additional spectral information, especially at lower frequencies, will be crucial for a more comprehensive understanding.

\subsection{X-ray properties}
\label{sect:xray_prop}

\begin{figure}
    \centering
    \includegraphics[scale=0.4]{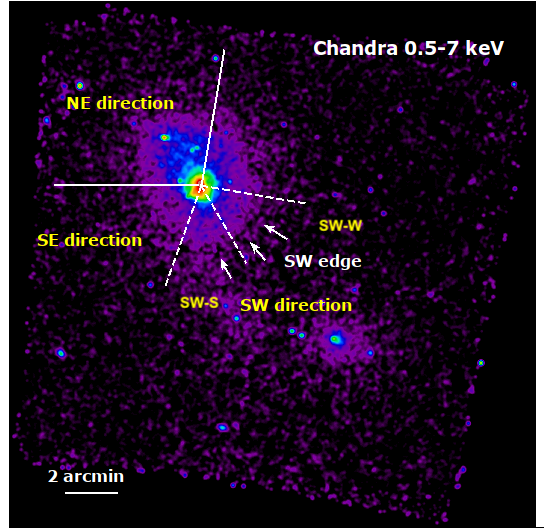}
    \caption{Selected directions for extracting the surface brightness and temperature of A3017. The SW direction is further divided into two equal angles.}
    \label{fig:sbp_reg}
\end{figure}

The exposure-corrected $0.5-2.0$~keV XMM-Newton images are presented in the right panel of Figure~\ref{fig:radio-xray}. The X-ray emission of A3017 exhibits an asymmetric and elongated morphology, including an X-ray bright core, X-ray excess toward the north, and a sharp surface brightness edge at $\sim 2\arcmin$ (i.e., $\sim 426$~kpc) southwest of the centre of A3017 (see also Figure~\ref{fig:sbp_reg}). An X-ray bridge containing a potential galaxy group is observed between A3017 and A3016 \citep{parekh17,chon19}.

\begin{figure}
    \centering
    \includegraphics[scale=0.28]{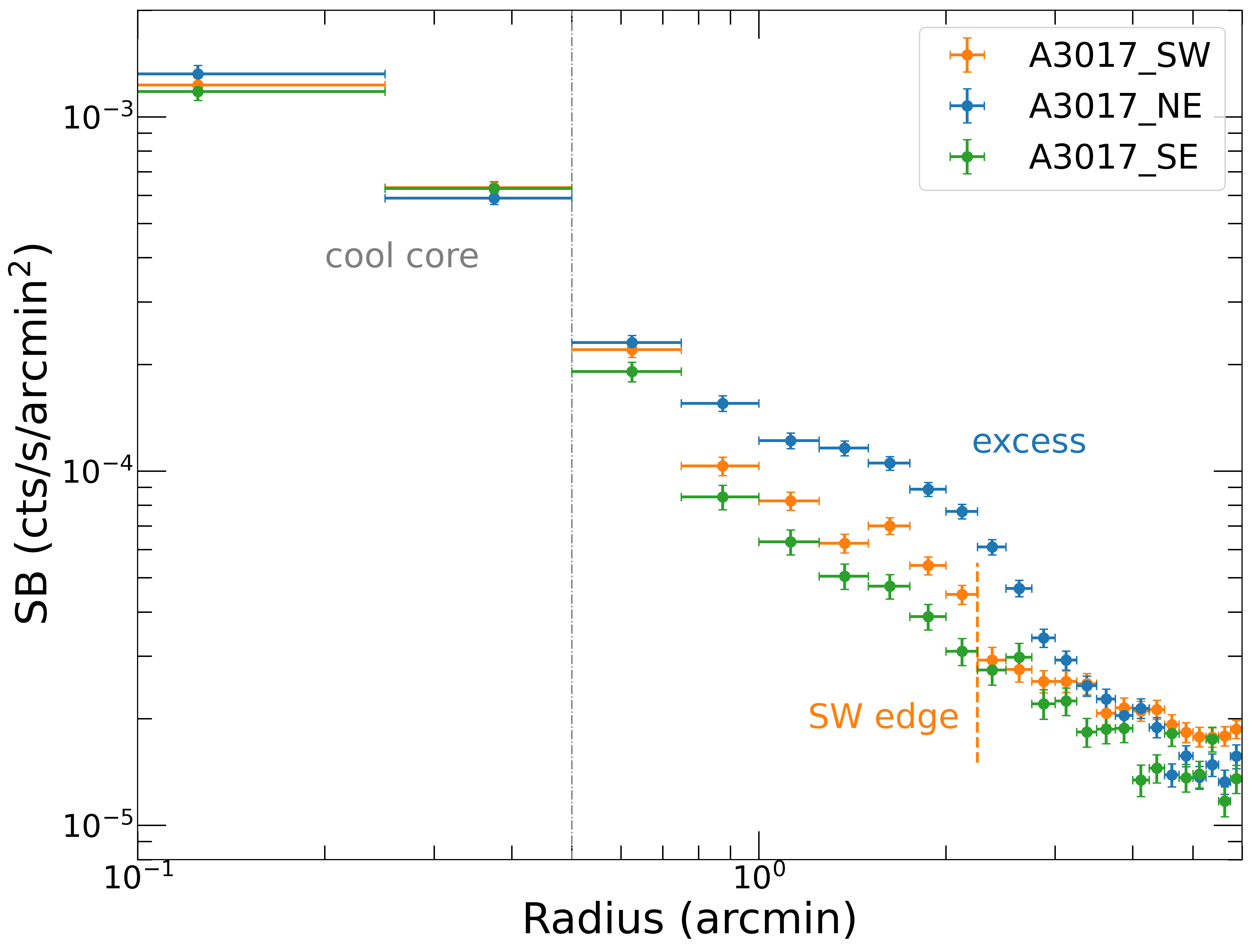}
    \caption{Surface brightness profiles derived in three directions. }
    \label{fig:sbp}
\end{figure}

\begin{figure*}
    \centering
    \includegraphics[scale=0.28]{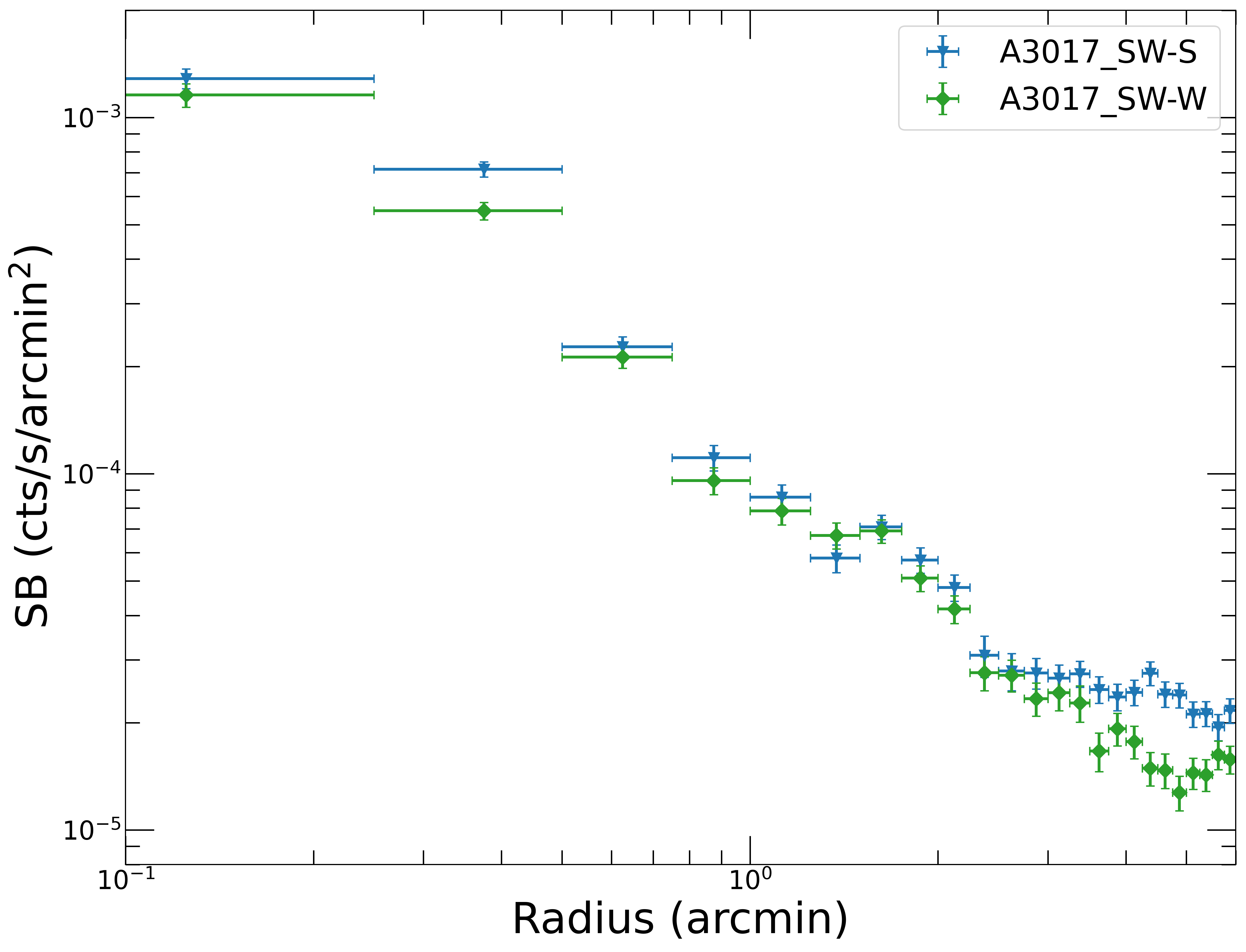}
    \includegraphics[scale=0.28]{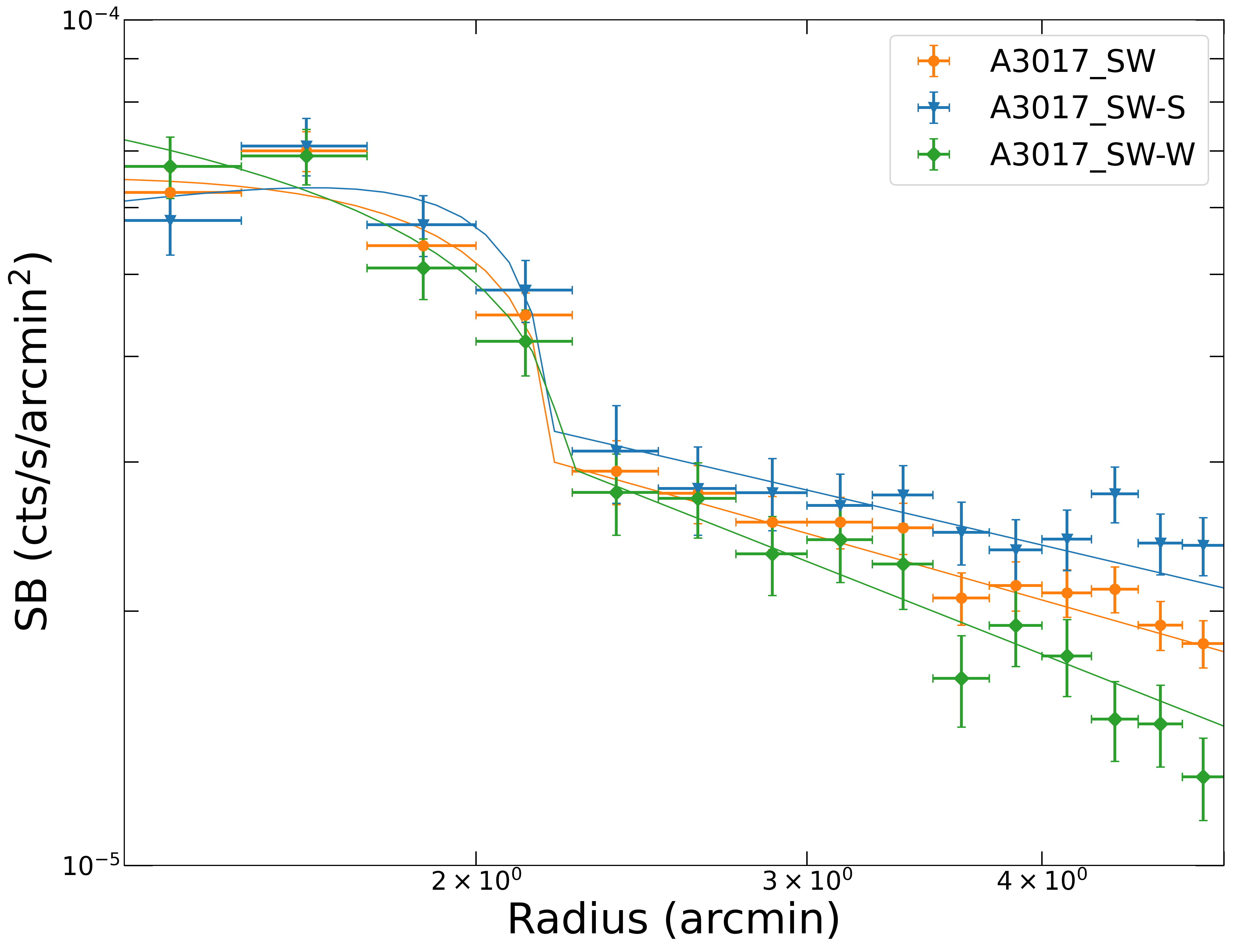}
    \caption{Left: Surface brightness profiles derived in two sub-sectors in the SW direction. Right: Surface brightness profiles in three SW directions with best-fit results using a broken power-law model. }
    \label{fig:sbp_sub}
\end{figure*}

\subsubsection{X-ray surface brightness distributions}
\label{sect:SBP}

To quantify the southwest (SW) edge and identify any potential substructures in the northeast (NE) direction, we extracted surface brightness profiles along these two directions using Chandra data due to it has higher spatial resolution. For comparison, we also present the surface brightness profile in the southeast (SE) direction. The extracted surface brightness profiles are presented in Figure~\ref{fig:sbp}. 

An inner surface brightness edge (with the most prominent one marked in the top-right and bottom-left panels of Figure~\ref{fig:a3017core}) is clearly shown in three profiles at $\sim 0\arcmin.5$ (i.e., $\sim 106$~kpc), which is corresponding to the edge of the cool core.  
To characterize the SW edge, the SW surface brightness profile was modelled using a projected three-dimensional density profile implemented in \texttt{PYPROFFIT} \citep{eckert20}, adopting a broken power-law form:
\begin{equation}
    n_{\rm e}(r) =
        \begin{cases}
            Cr^{-\alpha_{1}}  &  \text{if } r \leq r_{\rm break} \\
            C\frac{1}{d_{\rm jump}}r^{-\alpha_{2}} &  \text{if } r > r_{\rm break} ,
        \end{cases}
\end{equation}
Where $\alpha_{1}$ and $\alpha_{2}$ are the inner and outer slope indices at the break radius ${r_{\rm break}}$, and $d_{\rm jump} = n_{\rm e, in}/n_{\rm e, out}$ represents the density jump ratio between the inner ($n_{\rm e, in}$) and outer ($n_{\rm e, out}$) regions. 
The best fit indicates a density jump at $\sim 2\arcmin.20 \pm 0\arcmin.06$, with a ratio of $d_{\rm jump} = 2.20 \pm 0.4$. According to the Rankine-Hugoniot jump condition \citep{LL59}, the Mach number ($\mathcal{M}$) can be derived from the following equation:
\begin{align}
    d_{\rm jump} = \frac{4\mathcal{M}}{\mathcal{M}^{2} + 3} 
\end{align}
The X-ray derived Mach number based on the electron number density jump is $\mathcal{M}_{\rm X, n_e} = 1.91 \pm 0.21$.

To accurately estimate the surface brightness edge in the SW direction, we further divided the SW region into two equal parts. The eastern part (referred to as SW-S) is towards the direction covering the X-ray bridge, and another part (referred to as SW-W) avoids the bridge region. We applied the broken power-law model to both regions to determine the location of the edge and the density ratio. The profiles and fitted models are presented in Figure~\ref{fig:sbp_sub}. The best-fit results suggest that the edge is located at the $2\arcmin.18 \pm 0\arcmin.05$ and $2\arcmin.22 \pm 0\arcmin.11$, with the density jump ratios of $2.5 \pm 0.6$ and $1.79 \pm 0.35$ for the SW-S and SW-W directions, respectively. 
 
Compared to the surface brightness profile in the SE direction, the profile in the NE direction exhibits an apparent surface brightness excess at a radius of $\sim 1\arcmin - 3\arcmin$ (i.e., $213-639$~kpc), aligning with both the X-ray extension and also the N-extension in the radio image. The absence of a surface brightness jump associated with the N-ridge in both Chandra and XMM-Newton data could be attributed to either insufficient Chandra exposure time or the limited spatial resolution of XMM-Newton.

\subsubsection{X-ray spectral analysis}
\label{sect:temp}

\begin{figure}
    \centering
    \includegraphics[scale=0.28]{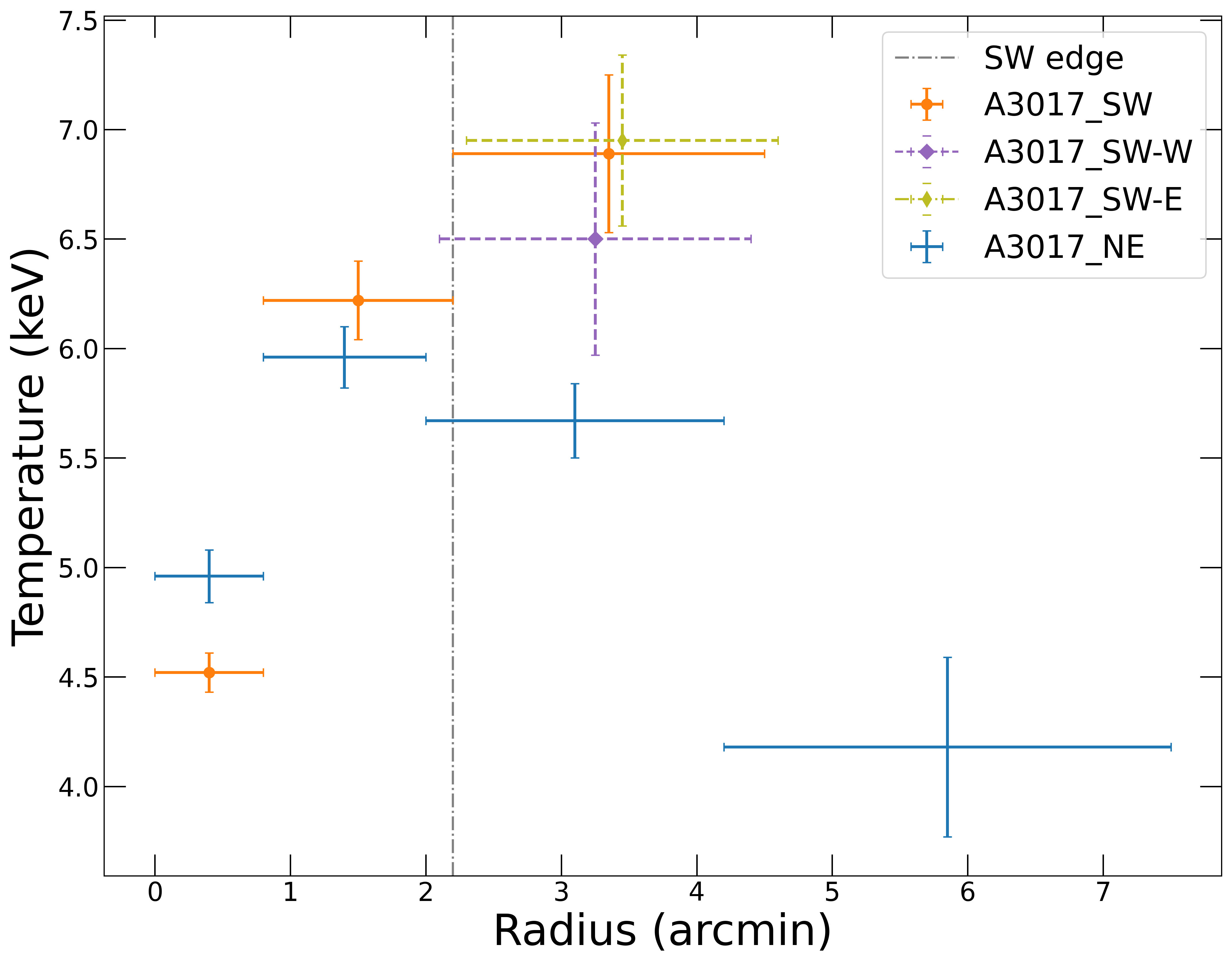}
    \caption{Gas temperature profiles derived along the NE and SW directions. Temperatures at the third annulus derived from two sub-sectors in the SW direction are also presented.}
    \label{fig:temp_prof}
\end{figure}

\begin{figure*}
    \centering
    \includegraphics[width=0.95\textwidth]{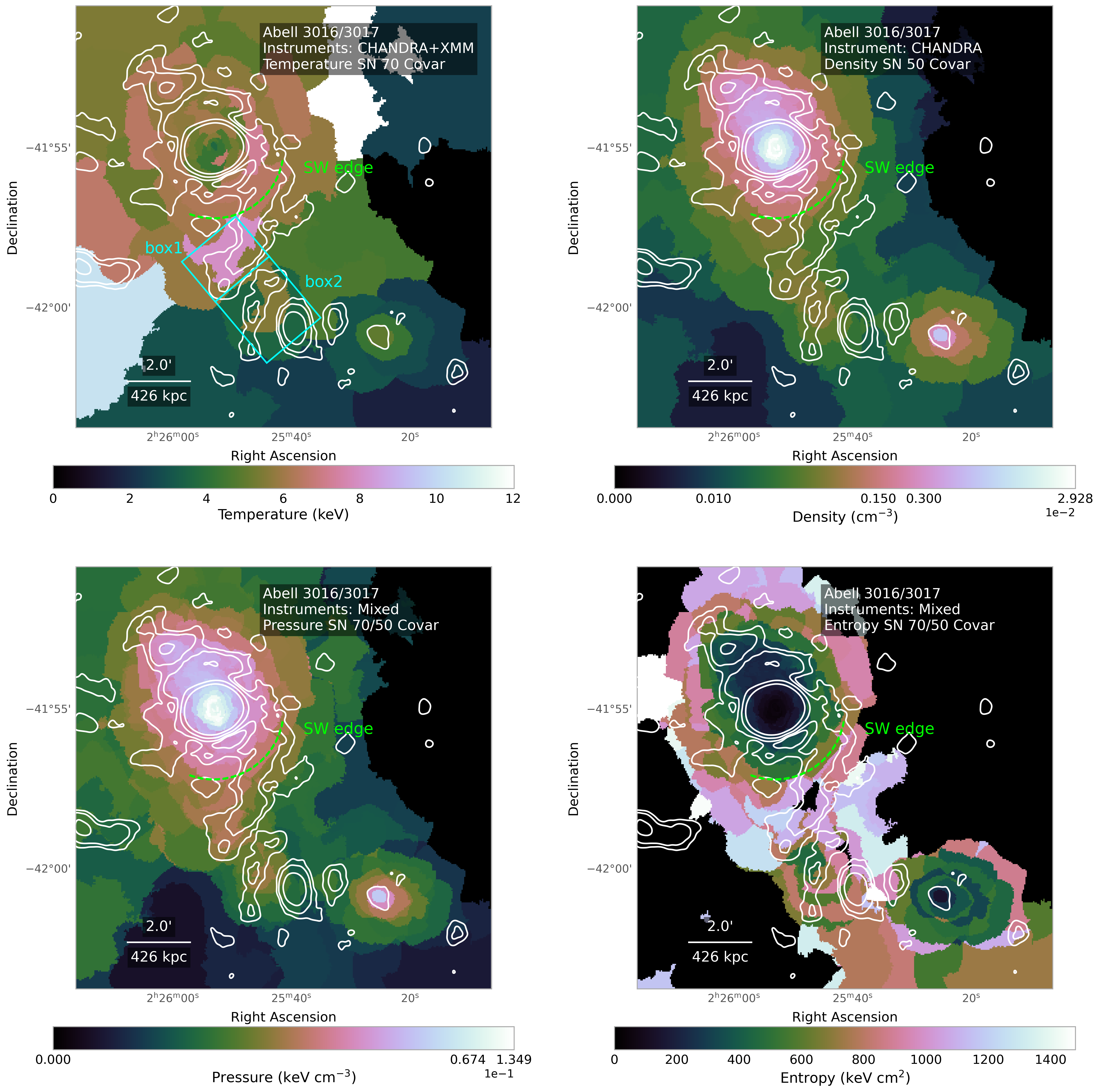}
    \caption{Two-dimensional projected thermodynamic maps of temperature (top-left), electron density (top-right), pressure (bottom-left), and entropy (bottom-right). The temperature map was created using binning with an S/N of 70, and the electron density map was created using binning with an S/N of 50. The pressure and entropy maps were created using the maps from temperature and electron density. The SW edge is marked, and the contours of the diffuse radio emission at high resolution are overlaid. Box regions used for estimating the temperature of the X-ray bridge are also indicated.}
    \label{fig:2dmap_covar}
\end{figure*}

\begin{figure*}
    \centering
    \includegraphics[width=0.95\textwidth]{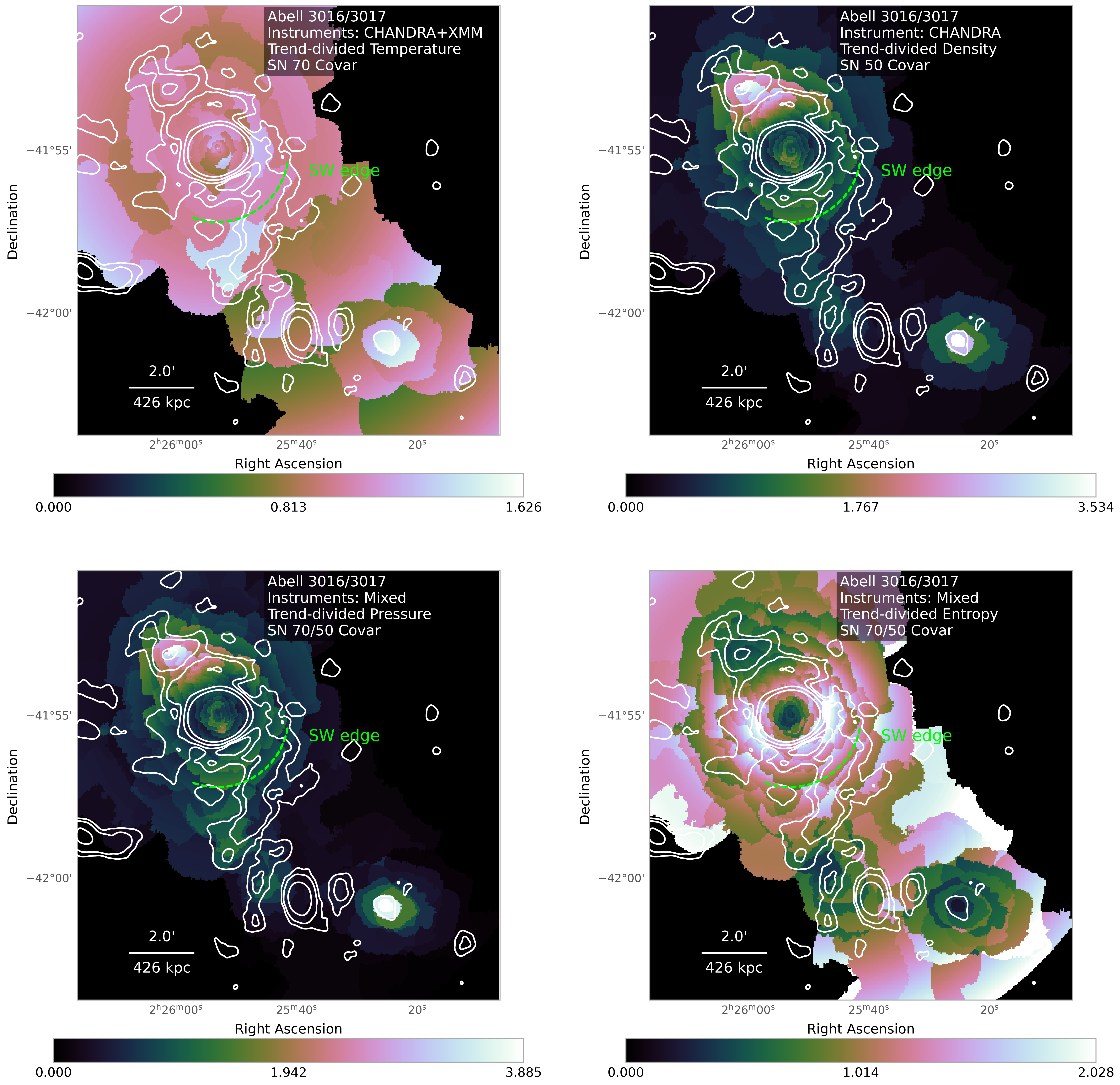}
    \caption{Azimuthal trend-divided maps, showing the deviations of temperature (top-left), electron density (top-right), pressure (bottom-left), and entropy (bottom-right) from radially averaged models centred on A3017. The SW edge is marked, and the contours of the diffuse radio emission at high resolution are overlaid.}
    \label{fig:trend-div}
\end{figure*}

To examine the potential shock front corresponding to the SW edge, as reported in \cite{chon19}, we extracted spectra from SW and NE directions using Chandra and XMM-Newton data and joint-fitted them by the model \texttt{tbabs*apec} to account for the ICM emission. The gas temperature profiles are shown in Figure~\ref{fig:temp_prof}.
At the location of the surface brightness jump $\sim 2\arcmin.2$ in the SW direction, no gas temperature drop is detected. This non-detection may be due to the complex dynamics of the system or the lack of consideration for projection effects, as only a single region outside the SW edge could be selected due to data quality constraints and the need to avoid contamination from A3016.

To have an overall understanding of A3016-A3017, we generated two-dimensional thermodynamic maps and azimuthal trend-divided maps using ROSAT, Chandra, and XMM-Newton data, and presented them in Figures~\ref{fig:2dmap_covar} and~\ref{fig:trend-div}, respectively. The modified version of the contour binning algorithm described in \citet{Sanders06} and \citet{breuer24} was used with the smoothed XMM-Newton images to produce polygon regions that could be independently extracted from the Chandra and XMM-Newton event lists. Before fitting these regions, the full CXB part of the spectral model was constrained using ROSAT and the entire FOV of Chandra and XMM-Newton, excluding the CCD edges, all point sources, and other emission regions. With the CXB constrained, each of these independent map regions was fit with a full instrumental background model corresponding to the specific area of the detector the region corresponds to described in Section~\ref{sect:xmm}, as well as the full CXB + Cluster model (corresponding to, \texttt{apec + tbabs(apec + apec + powerlaw) + tbabs(apec)}, also described in Section~\ref{sect:xmm}).

Due to available data quality, contour bin regions corresponding to 10,000, 4900, and 2500 counts per bin (Signal/Noise 100, 70, 50, respectively) were extracted from Chandra, and bins with 10000 and 4900 counts were extracted from XMM-Newton. Temperatures were derived spectroscopically for each region from the joint fit of Chandra and XMM-Newton spectra for S/N 70, while electron density ($n_\textrm{e}$) was calculated from Chandra spectra for S/N 50. The electron density was obtained from the normalization parameter ({\it norm}), defined as
\begin{equation}
\textit{norm} = \frac{10^{-14}}{4\pi (d_\textsc{a} (z+1))^2}\int n_\textrm{e} n_\textrm{p} dV \,,
\label{eq:norm}
\end{equation}
where $d_\textsc{a}$ is the angular diameter distance, $z$ is the redshift, and $\int n_\textrm{e} n_\textrm{p}$ is the integrated cluster emissivity over the chosen volume of gas in the cluster. The electron density $n_\textrm{e}$ and ion density $n_\textrm{p}$ are related as $n_\textrm{e} = 1.18 n_\textrm{p}$. Here, a constant line-of-sight depth of 1~Mpc is assumed for simplicity. The corresponding electron pressure and entropy maps were then calculated as $P_e = n_\textrm{e} kT$ and $K = kT n_\textrm{e}^{-2/3}$, respectively. Asymmetric errors were estimated by resampling the parameter distributions 100,000 times from the covariance matrix and the 68\% confidence intervals (corresponding to 1$\sigma$) were derived from the median, along with the 16th and 84th percentiles of the resampled distribution. Figure~\ref{fig:2dmap_covar} shows the temperature map (S/N 70) and electric density map (S/N 50), along with the derived pressure and entropy maps. Regions where the relative errors exceeded $3\sigma$ (i.e., 99.7\% confidence level) were blanked out.

The azimuthal trend-divided maps were then computed following the procedure described in \citet{ichinohe15} and \citet{breuer24}. In this approach, scatter plots of the radial distribution of each physical quantity from the A3017 cluster centre are generated, and the average trend is fit using a function of the form,
\begin{equation}
f(r) = A (1+ (r/B)^{2})^{(-3C/2)} (1+ (r/D)^{2})^{(-3E/2)}  \,,
\label{eq:trend_model}
\end{equation}
where $r$ is the radial distance from the centroid, and $A$, $B$, $C$, $D$, $E$ are free parameters. The residual after dividing the azimuthal average enhances the small variations present in the cluster, as seen in Figure~\ref{fig:trend-div}.

From the temperature map and its trend-divided map, the temperature jump according to the SW surface brightness edge is not apparent. However, indications of density and pressure discontinuities at the SW edge are presented in the density and pressure maps, as well as in their trend-divided maps, indicating the presence of a weak shock front.
An apparent density excess is observed north of A3017 in the trend-divided density map. At the same location, a pressure excess and an entropy deficit are also evident in the trend-divided maps of pressure and entropy. Given the absence of an obvious temperature jump in the northern part of A3017, this substructure is more likely to represent a subcluster.

In the temperature map (see Figure~\ref{fig:2dmap_covar}), we notice a high-temperature ($8.03 \pm 0.84$~keV) region to the south of the SW edge, located within the X-ray bridge that connects cluster A3017 to the potential galaxy group.
We selected two box regions along the X-ray bridge to represent the connecting region (box1) and the potential group region (box2), respectively. The gas temperature in box1 is $7.09 \pm 0.54$~keV, while in box2, the temperature is $4.21 \pm 0.27$~keV.
As mentioned in \cite{chon19}, the high temperature in the X-ray bridge region indicates gas heating due to the interaction, such as a shock or compression, between the two clusters. Additionally, the high temperature in the connecting region (box1) could suggest an additional interaction between A3017 and the potential embedded group.

\subsection{Correlations between thermal and non-thermal emission for diffuse radio structures}
\label{sect:Ir-Ix}

\begin{figure*}
    \centering
    \includegraphics[width=0.32\textwidth]{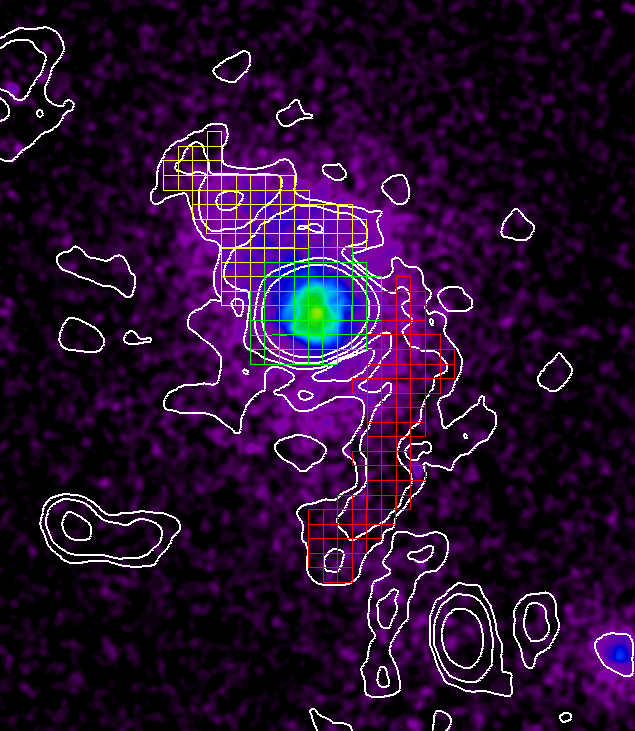}    \includegraphics[width=0.6\textwidth]{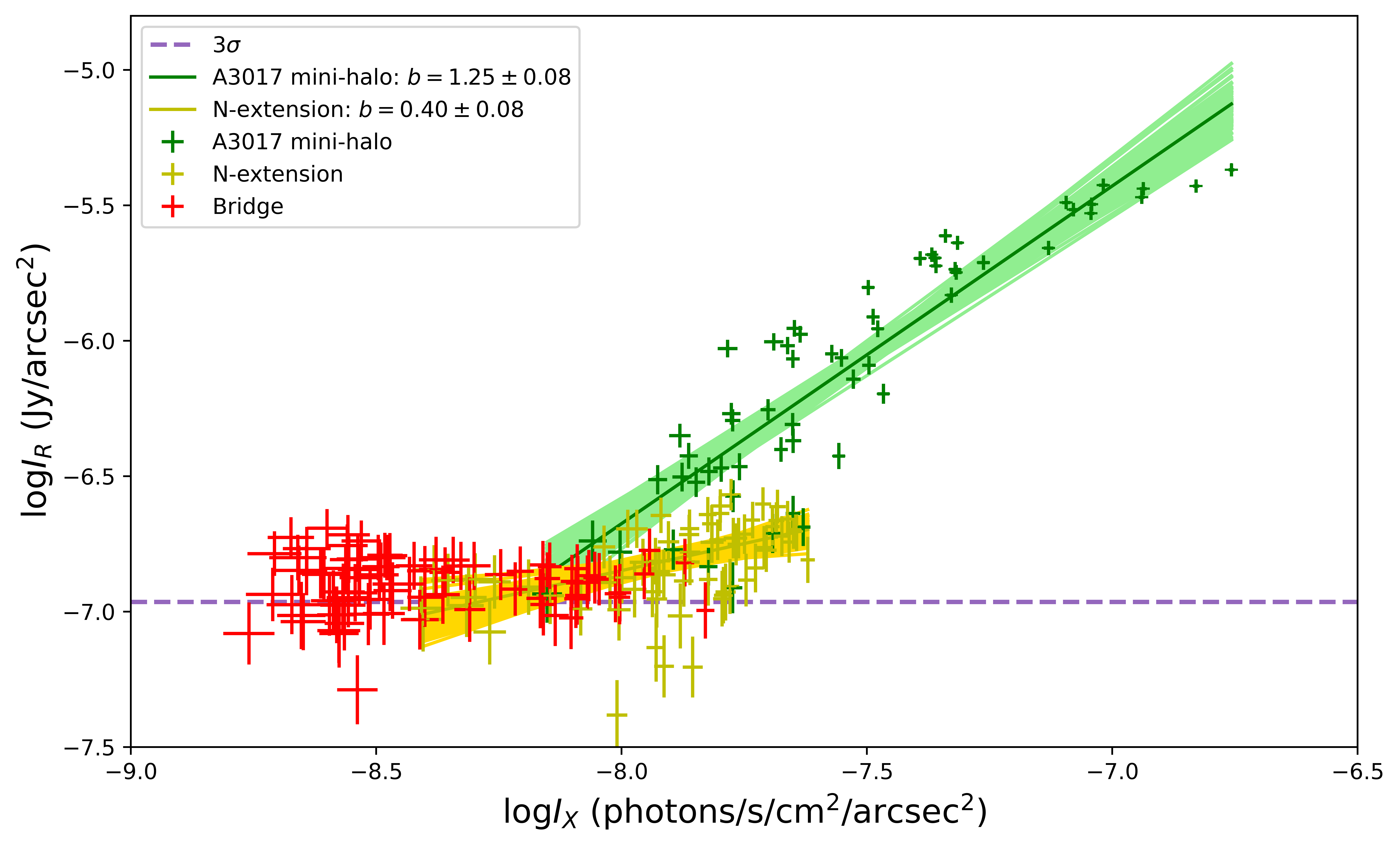}
    \caption{$I_{\rm R} - I_{\rm X}$ correlation of diffuse radio structures in A3016-3017. 
    Left: Point-source subtracted Chandra image overlaid with radio contours from compact-source-subtracted high-resolution  MeerKAT image after removing compact sources. The grid used to extract the surface brightness correlation is also presented. 
    Right: $I_{\rm R} - I_{\rm X}$ correlations for the A3017 mini-halo (in green), N-extension (in yellow), and bridge (in red). The values of X-ray surface brightness ($I_{\rm X}$) are extracted from the point-sources-subtracted XMM-Newton image. The horizontal dashed line indicates the $3\sigma$ in the compact-source-subtracted high-resolution MeerKAT image.}
    \label{fig:sb_grid}
\end{figure*}

The correlation between diffuse radio emission and the ICM, traced by X-ray emission, provides essential insights into the relationship between thermal and non-thermal components of galaxy clusters. The correlation between the radio ($I_{\rm R}$) and X-ray surface brightness ($I_{\rm X}$) on a point-to-point basis can help to explore the particle acceleration mechanisms responsible for diffuse radio emission, such as radio halo and radio mini-halo \citep{govoni01,riseley22,bonafede22,rajpurohit23}. 
For example, \cite{govoni01} and more recent investigations by \cite{botteon20b} and \cite{rajpurohit21}, have found that radio halos typically exhibit a sublinear or linear correlation, while mini-halos may display linear or superlinear scaling.
Additionally, quantitative comparisons of radio and X-ray surface brightness distributions further reveal spatial correlations, indicating the complex interplay between these components in different environments, e.g., inside and outside the cluster core \citep{biava21,rajpurohit23}.  

To investigate the thermal and non-thermal correlation of the diffuse radio structures in A3016-A3017, we performed a point-to-point analysis using the compact-source-subtracted, high-resolution MeerKAT 1280~MHz image and the point-source-subtracted XMM-Newton image. 
A grid with $15\arcsec$ square cells was applied to cover the diffuse radio structures, including the A3017 mini-halo, N-extension, and radio bridge. The selected grid and resulting $I_{\rm R}-I_{\rm X}$ correlation are shown in Figure~\ref{fig:sb_grid}. 
A positive correlation between X-ray and radio surface brightness is observed for the A3017 mini-halo and N-extension. The relationship can be generally described by a power law:
\begin{align}
    {\rm log}(I_{\rm R}) = a + b~{\rm log}(I_{\rm X})
\end{align}
where $b$ is the correlation slope. When $b>1$ (i.e., superlinear correlation), it suggests that the radio emission related to magnetic field strength and relativistic electron density declines faster than the X-ray emission, which is associated with thermal electron density and gas temperature, or vice versa (if $b<1$).  

To quantify any potential correlations of diffuse radio structures of A3016-A3017 and consider uncertainties for both radio and X-ray surface brightness, we used \texttt{Linmix} \citep{kelly07} to fit the derived results of A3017 mini-halo and N-extension.  
The superlinear slope ($b = 1.25 \pm 0.08$) of the A3017 mini-halo is consistent with those cases of radio mini-halos \citep{biava21,riseley22}. 
In contrast, the sublinear slope ($b = 0.40\pm 0.08$) of the N-extension possibly indicates a different turbulent motion as the N-extension might be related to the infalling subcluster. The trend of the superlinear slope in the halo core and sublinear in the outer fainter region is also reported in the galaxy clusters A2256 \citep{rajpurohit23} and A1413 \citep{lusetti24} at LOFAR 144~MHz. 
However, the $I_{\rm R}-I_{\rm X}$ relation for the bridge is offset from the trend of the mini-halo and N-extension, suggesting a different connection between thermal and non-thermal components within the bridge region. 
Note that, the slopes of the $I_{\rm R}-I_{\rm X}$ relations for the N-extension and the bridge may be affected by the radio surface brightness limitations, such as the $3\sigma$ threshold indicated in Figure~\ref{fig:sb_grid}. Future deeper radio observations at lower frequencies are needed to further investigate these two substructures.

\section{Discussion}
\label{sect:discussion} 

\subsection{Radio bridge}
\label{sect:radio-bridge} 

The radio bridge between the two clusters is clearly detected in both high- and low-resolution MeerKAT images after subtracting compact sources (Figure~\ref{fig:radio-subbands}). The average surface brightness of the radio bridge at 1280~MHz is $\sim 0.1~\rm \mu Jy~arcsec^{-2}$ (derived from the compact-source-subtracted, high-resolution MeerKAT image, refer to red dots in Figure~\ref{fig:sb_grid}), which is too faint to be detected by previous telescopes. This radio bridge connects cluster A3017 with a potential galaxy group located within the bridge and extends toward cluster A3016, spanning $\sim 5\arcmin.6$ (i.e., $\sim 1.2$~Mpc) in length, in projection. This radio bridge aligns well with the X-ray bridge (Figure~\ref{fig:radio-xray}).
Based on the optical evidence provided by \cite{foex17}, which indicates that (1) the A3017 cluster resides in a rich large-scale environment, (2) the A3016 cluster and the galaxy group are also located along this large-scale accretion axis and are likely to merge into A3017, and (3) these three systems are very likely gravitationally bound, with a merger plane close to the plane of the sky, we propose three possible scenarios for the origin of the radio bridge.

\noindent\textit{1. Inter-cluster radio bridge in the filament:}\\
First, the radio bridge could represent an inter-cluster radio bridge between two pre-merger clusters A3016 and A3017, similar to those discovered between A399-A401 \citep{govoni19} and A1758 \citep{botteon20c}. 
As the two clusters approach each other, interactions between matter in their outskirts generate turbulence and shock waves. These interactions could amplify magnetic fields and re-accelerate relativistic electrons, resulting in diffuse radio emission that appears as a bridge between the clusters. Additionally, the presence of a potential galaxy group, as a small-scale concentration of galaxies and matter within the filament between the two clusters, likely contributes to dynamical activity in this region, as evidenced by the high temperature observed in the X-ray bridge. The movement of this group within the filament could also induce turbulence and shocks, re-accelerating fossil radio plasma and gently sustaining relativistic electrons over long timescales.
However, unlike the cases observed in A399-A401 and A1758, the radio bridge in A3016-A3017 extends towards A3016 but does not fully connect to its radio emission, and no radio halo is detected in A3016. This could be attributed to the relatively modest total mass of A3016, estimated as $M_{\rm 500}^{\rm A3016} \approx 1.3 \times 10^{14}$~$\rm M_{\sun}$ \citep{chon19}, which limits larger-scale accretion motions. As a result, neither a radio halo nor a fully extended radio bridge is detected on the A3016 side. Alternatively, the radio emission from the A3016 side may simply be too weak, and the radio halo and connecting bridge could only be visible at lower frequencies. 

\noindent\textit{2. Radio bridge results from interactions between A3017 and the group:} \\
An alternative possibility is that the radio bridge results from interactions between A3017 and the galaxy group, similar to those detected between the Coma cluster and the NGC~4839 group (\citealt{bonafede21}; see also \citealt{lyskova2019,sheardown2019,churazov2021}), as well as A3562 and the SC1329-313 group \citep{venturi22}. In this scenario, the galaxy group is undergoing an off-axis merging process with A3017. It has passed through its primary apocenter and is now falling back towards A3017. The merger shock front formed before the apocentric passage continues propagating towards the southwest \citep{czhang2019}. The post-shock turbulence may sustain the re-acceleration of relativistic electrons, leading to diffuse radio emission extending beyond the group towards A3016. As the galaxy group turns back for its second infall towards A3017, the newly formed bow shock formed in front of the group may explain the high-temperature region detected between the A3017 and group (i.e., box1; see Section~\ref{sect:temp} and Figure~\ref{fig:2dmap_covar}).
However, based on the current data of A3016-A3017, we are unable to determine the dynamic status of the group associated with A3016 and A3017. Although the observed $I_{\rm R}-I_{\rm X}$ relation of the bridge indicates an anticorrelation, this relationship may be affected by the $3\sigma$ radio detection limitations.  
We found that the A3016-A3017 system closely resembles the pre-merger system A2061-A2067 recently observed at LOFAR 144~MHz \citep{pignataro24}. In the A2061-2067 system, the diffuse radio emission (i.e., NE extension) was detected connecting the radio halo of A2061 and extending toward A2067, where a radio halo is also absent. Given an X-ray plume in A2061 is likely related to the potential sub-group between two clusters, \cite{pignataro24} proposed that (1) if the sub-group is falling into A2061 from the NE, creating an X-ray plume as the galaxies precede their intracluster medium, and the NE extension could represent a radio bridge; (2) if the sub-group has already passed through the A2061 core and is now returning for a second infall, the X-ray plume is formed as its slingshot gas tail and the NE extension as an elongation of the radio halo. 
Therefore, the dynamics of the galaxy group play a crucial role in understanding the true origin of the radio bridge, as pointed out by \cite{pignataro24}.

\noindent\textit{3. Cluster radio relic interpretation:}\\
In addition to the above two possibilities, the radio bridge might also represent a cluster radio relic associated with a cluster merger shock, causing the thermal electrons or fossil relativistic electrons to be (re-)accelerated to emit radio emission. 
If this is indeed a merger-induced radio relic, we would expect to detect a shock front ahead of the relic. However, possibly due to the limitations of the current X-ray data quality, we are unable to detect this outer shock front, either in terms of surface brightness or temperature.
Additionally, the radio surface brightness within this diffuse radio emission does not show a typical decline toward the cluster centre, which is a characteristic commonly observed in cluster radio relics. This might indicate we are observing a radio relic from a face-on perspective. The TNG-Cluster simulation suggests that `face-on' radio relics can have a morphology similar to radio halos, but do not align with the discontinuities seen in X-ray emission \citep{lee24}. However, considering the X-ray morphology of the A3017 system and the optical evidence indicating that the large-scale filament is nearly parallel to the plane of the sky, it seems unlikely that the merger axis in A3017 is aligned along our line of sight.
Due to the limitations in current data quality, we are unable to investigate the spectral index distribution thoroughly. Future deep observations and polarization analysis are needed to examine the possibility of the cluster radio relic.

Another important consideration is the origin of the seed electrons. There is no prominent radio galaxy within the region of the radio bridge to supply the fossil seed electrons. One possible explanation is that, as the shock front passed, thermal electrons were accelerated according to the diffusive shock acceleration (DSA) model. The radio power of the radio bridge can be estimated as $L_{1.4~\rm GHz} \approx 4.15 \times 10^{23}$~$\rm W~Hz^{-1}$ by adopting a spectral index of $\alpha_{\rm fit}^{\rm bridge} = -2.13$. This steep spectral index leading to the Mach number of $\mathcal{M}_{\rm radio} \approx 1.66$ suggests that the shock front here is probably inefficient to produce the observed radio luminosity \citep{botteon20a,botteon24}. 
Therefore, the most likely explanation is the presence of in situ relativistic fossil electrons or mildly relativistic electrons between the two clusters \citep{govoni19,botteon20c}. 
Meanwhile, the relativistic electrons could also be attributed to the galaxy group. When the group moves through the triple system, the inter-group medium may be stripped away, carrying with it relativistic electrons supplied by the AGN within the group.

\subsection{Extended radio emission in A3017}
\label{sect:radio-halo} 

The MeerKAT 1280~MHz image reveals that the diffuse radio emission in A3017 contains two different components: a central bright emission (i.e., radio mini-halo) and an outer diffuse radio emission (e.g., N-extension). This is further supported by the radio surface brightness profile, suggesting a steep inner component and a flatter, fainter outer component.
The central radio emission spans $\sim 277$~kpc at $9\arcsec$ resolution (Figure~\ref{fig:a3017core}), similar to the typical size range of $100-500$~kpc for radio mini-halos \citep{vanweeren19}. It is located in the cluster's bright cool core, where the temperature is relatively low ($\sim 4.5$~keV, see Figure~\ref{fig:temp_prof} and ~\ref{fig:2dmap_covar}; \citealt{parekh17,pandge21}), and spatially corresponds to but is bound within the inner X-ray surface brightness edge (see in Figure~\ref{fig:a3017core}). 
Given the size and location of the central radio emission, and the presence of a central AGN with two lobes, the central emission likely represents a radio mini-halo.
The outer diffuse radio emission shows a northern extension, i.e., N-extension, aligning with the northern X-ray excess of A3017. It could represent the outer radio halo, spanning roughly $\sim 1$~Mpc and mostly filling the ICM of A3017.
Additionally, the correlation between $I_{\rm R}$ and $I_{\rm X}$ suggests a superlinear slope for the mini-halo and a sublinear slope for the N-extension, implying different mechanisms may be at play for these two radio components. 

The co-existence of radio mini-halos and giant radio halos has increasingly been detected, mainly by LOFAR at low frequencies \citep{bonafede14,venturi17,savini18,biava21,lusetti24,vanweeren24}. Most of these detections occur in cool-core clusters, a common environment for radio mini-halos. The discovery of outer radio halos raises questions about the long-held view that giant radio halos are typically found in disturbed clusters.
For example, \cite{vanweeren24} recently discovered a giant radio halo, with a projected size of $1.1$~Mpc, beyond the mini-halo in the Perseus Cluster. The Perseus Cluster is notable for its multiple detected cold fronts, with the farthest confirmed one located 0.7~Mpc from the cluster core \citep{simionescu12,walker18,bellomi2024}. This cluster experienced an offset merger at least 5~Gyr ago, and the large-scale gas sloshing and turbulence induced by this merger are thought to provide sufficient energy to power the outer radio halo. \cite{biava24} studied 12 cool-core clusters and found that outer radio halos were not detected in all clusters at LOFAR 144~MHz. However, when detected, these outer halos were often associated with cold fronts. 

In contrast, A3017 is a cool core cluster, but previous studies suggest it is still undergoing a merger process \citep{foex17,chon19}. The northern substructure, corresponding to the northern X-ray excess and radio N-extension, probably indicates a significant mass of infalling materials, estimated at $> 8 \times 10^{13} ~\rm M_{\odot}$ \citep{chon19}. 
In this case, turbulence is primarily driven by the infalling subcluster rather than the large-scale gas sloshing. This may explain the non-constraint of the outer component when fitting the radio surface brightness profile with a double exponential model (see Sector~\ref{sect:radio_image}). The outer radio emission may have an additional and possibly more recent contribution from the infalling subclusters, in addition to the gas sloshing.
The presence of the SW edge, a potential shock front, suggests that at least part of the merger between A3017 and the infalling subcluster is progressing in this direction. 
Therefore, we propose that the subcluster may have completed its first passage in an offset merger with A3017, leaving the cool core of A3017 intact. This offset merger has generated gas sloshing in the cluster core, and the resulting turbulence, possibly in combination with interactions between AGN jets and the ICM, re-accelerated relativistic electrons injected by the central AGN, forming the mini-halo. 
Currently, the subcluster is either away from the cluster core or approaching the cluster core from the north for a second pericenter passage, likely generating large-scale turbulence along its path, which re-accelerates radio particles and contributes to the outer radio halo.

The results from A3017 support the view that the definition of the radio halo should take into account the specific merger status of the cluster, rather than simply distinguishing between merging or relaxed clusters, as suggested by \cite{vanweeren24}.  
The co-existence of radio mini-halos and giant radio halos may be a common phenomenon and is likely to be detected more frequently with the advent of advanced radio telescopes, such as LOFAR, MeerKAT, and SKA.

\subsection{Tadpole-like radio structure}
\label{sect:radio-vertical} 

\begin{figure*}
    \centering
    \includegraphics[scale=0.46]{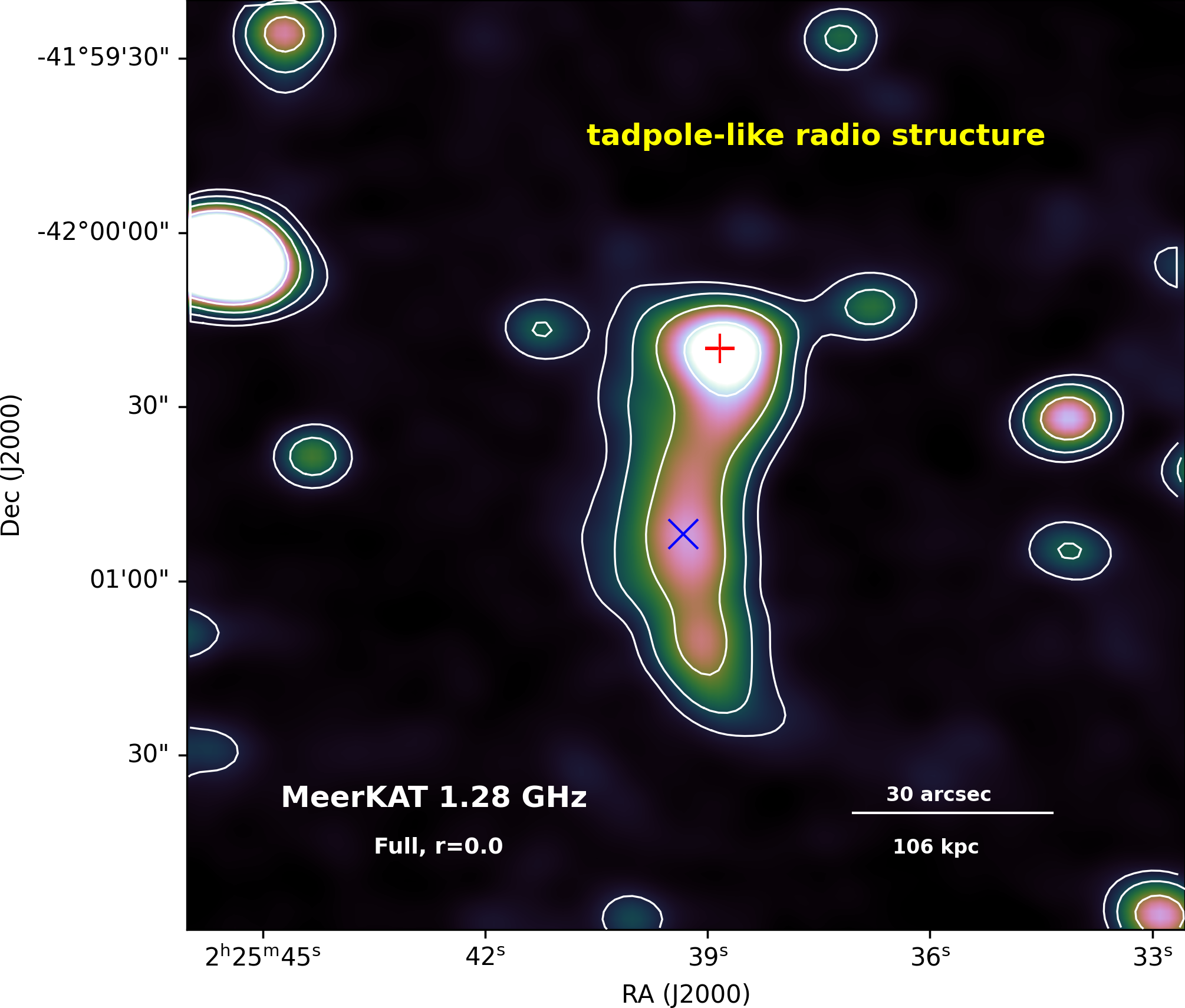}
    \hspace{0.2cm}
    \includegraphics[scale=0.46]{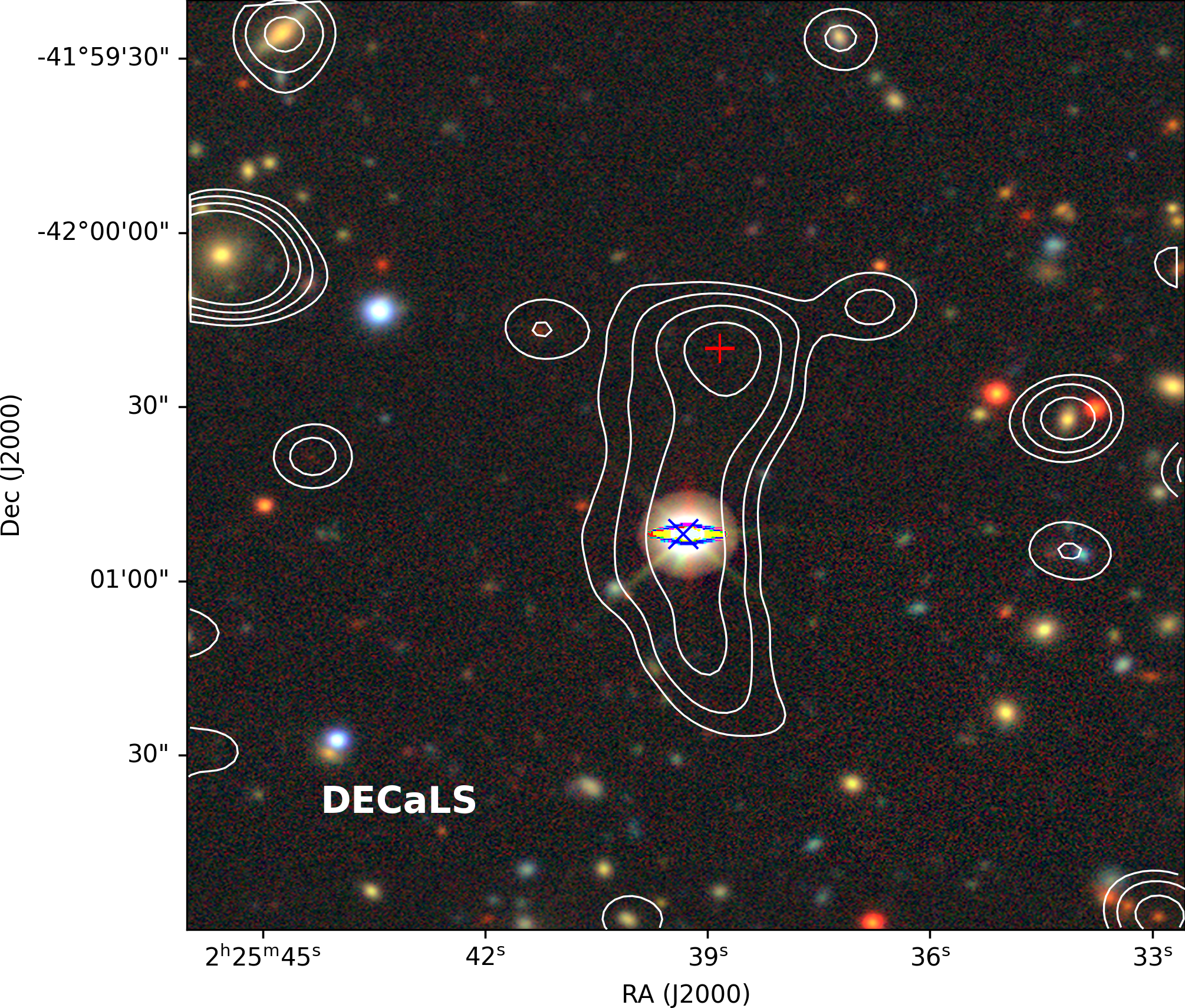}
    \caption{Left: Full-resolution MeerKAT 1280~MHz image of the tadpole-like radio structure with contour levels of [1, 2, 4, 8] $\times 5\sigma$. 
    Right: Composite DECalS optical image. The two radio bright spots are marked with "+" and "x".}
    \label{fig:vertical_src}
\end{figure*}

\begin{figure}
    \centering
    \includegraphics[scale=0.35]{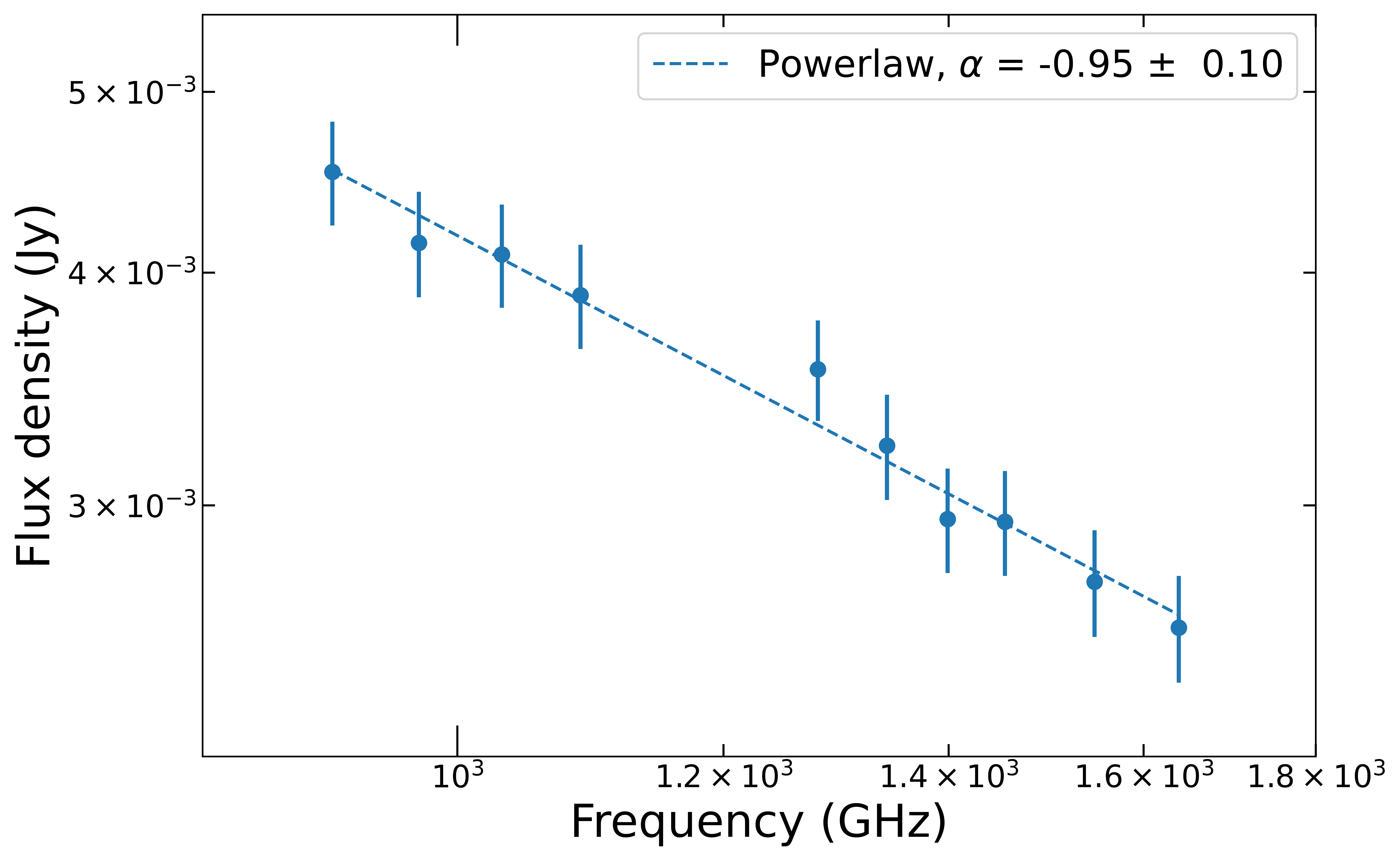}
    \caption{Radio spectrum of the tadpole-like radio structure.}
    \label{fig:vertical_src_spc}
\end{figure}

At the radio bridge between two clusters, there is a tadpole-like radio structure with an elongated morphology (Figure~\ref{fig:vertical_src}). This tadpole-like radio structure was first identified by \cite{parekh17} utilizing GMRT observations at 235 MHz and 610 MHz. Due to the lack of an optical counterpart or a corresponding compact radio source at other frequencies, they suggested that a possible weak shock within the bridge could be accelerating fossil relativistic electrons. However, the reported flat spectral index of $\sim -0.3$ is unusual, as such emission typically exhibits a much steeper spectral index ($\alpha < -1$; \citealt{vanweeren19}), raising questions about the true nature of this source. 

Owing to the better sensitivity of MeerKAT, the full-resolution MeerKAT 1280~MHz image clearly reveals two bright spots, which are marked in Figure~\ref{fig:vertical_src}. The upper, brighter one has no visible optical counterpart in the DECaLS image, while the lower, fainter one could potentially be linked to an optical source. This bright optical source could be a foreground star or a distant quasar. Based on the Gaia Early Data Release 3 (EDR3; \citealt{gaiacolla21})\footnote{Gaia EDR3 source ID: 4950015852354095488, with coordinates of RA, Dec: 02h25m39.2428s, -42d00m51.992s.}, this optical source has a proper motion of $\sim 7.9~\rm mas~yr^{-1}$ and a radial velocity of $\sim -28.8~\rm km~s^{-1}$, suggesting that it is likely a foreground star. Additionally, the absence of a corresponding X-ray point source rules out the possibility of it being a quasar. 

To investigate the spectral properties, we divided the MeerKAT L-band into 10 sub-bands and convolved them into the same resolution ($13\arcsec \times 13\arcsec$). The flux densities of this tadpole-like radio structure are presented in Figure~\ref{fig:vertical_src_spc}, and a single power-law fit suggests a spectral index of $\alpha = -0.95 \pm 0.10$.  
The spectrum of this radio structure may be dominated by the upper bright spot, which could correspond to a distant AGN. The elongated faint emission might represent the radio lobes or tails associated with this distant AGN or possibly multiple AGNs, leading to a steepening of the spectrum beyond typical values for AGNs or compact radio sources.
Alternatively, the entire radio structure could represent a radio relic within the radio bridge, possibly originating from a shock front induced by the motion of the galaxy group. In this scenario, the upper bright spot could either be the most luminous part of the relic due to the shock, or it might correspond to a distant AGN.

\section{Conclusions}
\label{sect:conclusion} 

In this paper, we present a composite radio and X-ray analysis of the pre-merger galaxy clusters A3016 and A3017. The main results are summarized as follows:
\begin{enumerate}
    \item The MeerKAT 1280~MHz image reveals a radio bridge connecting A3017 with a potential galaxy group and extending toward A3016. The radio bridge aligns with the X-ray bridge. It spans $\sim 1.2$~Mpc in projection and has an average surface brightness of $\sim 0.1$~$\rm \mu Jy~arcsec^{-2}$, making it too faint to have been detected previously. The in-band spectral index of the radio bridge is estimated as $\alpha_{\rm fit}^{\rm bridge} = -2.13\pm 0.52$. 
    \item The extended radio emission in A3017 consists of two components: a central radio mini-halo and an outer diffuse radio emission with a northern extension (i.e., N-extension), aligning with the northern X-ray excess of A3017. The mini-halo, with a size of $\sim 277$~kpc at $\sim 9\arcsec$ resolution, aligns with the X-ray core. The outer diffuse radio emission could represent the outer radio halo, spanning roughly $\sim 1$~Mpc and mostly filling the ICM of A3017.
    \item The radio surface brightness profile of the extended radio emission in A3017 reveals a steep inner component with an $e$-folding radius of $r_{\rm e, inner} = 21\arcsec.9 \pm 1\arcsec.1$ (i.e., $78 \pm 4$~kpc), consistent with the values found in other radio mini-halos. The flatter outer component has a large, poorly constrained $r_{\rm e, outer}$, which may be linked to the infalling subcluster, possibly associated with the N-extension. 
    \item The SW surface brightness edge, indicative of a possible shock front, was confirmed with a density jump at $\sim 2\arcmin.2$. However, no significant temperature drop was detected, likely due to the complex dynamics of the system or lack of consideration for projection effects.
    \item X-ray bridge exhibits a gas temperature of $4.21 \pm 0.27$~keV, suggesting gas heating due to the interaction between two clusters. Additionally, a high temperature of $7.09 \pm 0.54$~keV was detected in the connecting region in the bridge, suggesting gas heating from shocks or compression due to interactions between A3017 and the potential galaxy group. This may have re-accelerated relativistic electrons in the radio bridge. 
    \item The correlation between $I_{\rm R}$ and $I_{\rm X}$ reveals a superlinear slope for the A3017 mini-halo and a sublinear slope for the N-extension, indicating different underlying mechanisms for these two components.
    \item The tadpole-like radio structure in the bridge shows two bright spots, with the upper bright one lacking an optical counterpart and the lower one possibly linked to a foreground star. The spectral index of $\alpha = -0.95 \pm 0.10$ suggests the emission is likely dominated by a distant AGN or multiple AGNs, and the elongated faint emission could represent associated radio lobes or tails. Alternatively, this radio structure could represent a radio relic within the bridge generated by the shock front induced by the motion of the galaxy group.
\end{enumerate}

Based on the results, we propose three plausible explanations for the origin of the radio bridge: (1) it is an inter-cluster radio bridge connecting the two clusters in a filament, formed by the interaction between the two clusters and enhanced by additional interactions with the potential galaxy group located between them; (2) the radio bridge results from the interaction between A3017 and the galaxy group after their primary apocentric passage, with the group currently falling back towards Abell~3017; (3) it is a cluster radio relic associated with a merger shock, appearing as a bridge due to its face-on orientation.  
The absence of prominent radio galaxies within the bridge suggests that the relativistic electrons may come from in situ fossil electrons injected by the previous activities of AGNs or star formations. The galaxy group may also contribute relativistic electrons through its movement in this triple system. 
A3017 likely experienced an offset merger with the subcluster, leaving its cool core intact. The subcluster appears to have completed its first pericenter passage and is now either away from A3017 or approaching A3017 from the north, generating large-scale turbulence. The A3017 mini-halo is likely powered by the gas sloshing, with the relativistic electrons injected by the central AGN. The outer radio halo seems to be generated by turbulence primarily induced by the infalling subcluster.

\begin{acknowledgements}
We thank the referee for providing valuable comments and
suggestions.
DH, NW, JPB, CYZ, and TP acknowledge the financial support of the GA\v{C}R EXPRO grant No. 21-13491X. 
HGX, YYZ, HS and ZZL acknowledge the support of the National Natural Science Foundation of China (NFSC) Grant No. 12233005.
QZ, HYS, GQ, ZZH, and GJH acknowledge the support from the National SKA Program of China (no. 2020SKA0110100, 2020SKA0110200).\\

This work makes use of the MeerKAT telescope, operated by the South African Radio Astronomy Observatory, which is a facility of the National Research Foundation, an agency of the Department of Science and Innovation.
MeerKAT data published here have been reduced using the CARACal pipeline, partially supported by ERC Starting grant number 679627 “FORNAX”, MAECI Grant Number ZA18GR02, DST-NRF Grant Number 113121 as part of the ISARP Joint Research Scheme, and BMBF project 05A17PC2 for D-MeerKAT. Information about CARACal can be obtained online under the URL: https://caracal.readthedocs.io.\\

This research has made use of data obtained from the Chandra Data Archive provided by the Chandra X-ray Center (CXC), and observations obtained with XMM-Newton, an ESA science mission with instruments and contributions directly funded by ESA Member States and NASA. \\

This scientific work makes use of data obtained from Inyarrimanha Ilgari Bundara / the Murchison Radio-astronomy Observatory, operated by CSIRO. We acknowledge the Wajarri Yamatji people as the traditional owners of the Observatory site. Support for the operation of the MWA is provided by the Australian Government (NCRIS), under a contract to Curtin University administered by Astronomy Australia Limited. We acknowledge the Pawsey Supercomputing Centre which is supported by the Western Australian and Australian Governments. \\

The DESI Legacy Imaging Surveys consist of three individual and complementary projects: the Dark Energy Camera Legacy Survey (DECaLS), the Beijing-Arizona Sky Survey (BASS), and the Mayall z-band Legacy Survey (MzLS). DECaLS, BASS and MzLS together include data obtained, respectively, at the Blanco telescope, Cerro Tololo Inter-American Observatory, NSF’s NOIRLab; the Bok telescope, Steward Observatory, University of Arizona; and the Mayall telescope, Kitt Peak National Observatory, NOIRLab. NOIRLab is operated by the Association of Universities for Research in Astronomy (AURA) under a cooperative agreement with the National Science Foundation. Pipeline processing and analyses of the data were supported by NOIRLab and the Lawrence Berkeley National Laboratory (LBNL). Legacy Surveys also uses data products from the Near-Earth Object Wide-field Infrared Survey Explorer (NEOWISE), a project of the Jet Propulsion Laboratory/California Institute of Technology, funded by the National Aeronautics and Space Administration. Legacy Surveys was supported by: the Director, Office of Science, Office of High Energy Physics of the U.S. Department of Energy; the National Energy Research Scientific Computing Center, a DOE Office of Science User Facility; the U.S. National Science Foundation, Division of Astronomical Sciences; the National Astronomical Observatories of China, the Chinese Academy of Sciences and the Chinese National Natural Science Foundation. LBNL is managed by the Regents of the University of California under contract to the U.S. Department of Energy. The complete acknowledgements can be found at https://www.legacysurvey.org/acknowledgment/. \\

This work has made use of SAOImageDS9, CARTA \citep{comrie21}, and the following \textsc{python} packages: \texttt{astropy} \citep{astropy22}, \texttt{numpy} \citep{numpy11}, \texttt{matplotlib} \citep{matplotlib07}, and \texttt{scipy} \citep{scipy20}.

\end{acknowledgements}




%
   \bibliographystyle{aa} 
   \bibliography{reference} 


\begin{appendix} 

\onecolumn 
\clearpage

\section{MWA-2 images across five frequencies}

\vspace{0.5cm}

\begin{figure}[h]
    \centering
    \includegraphics[width=0.95\textwidth]{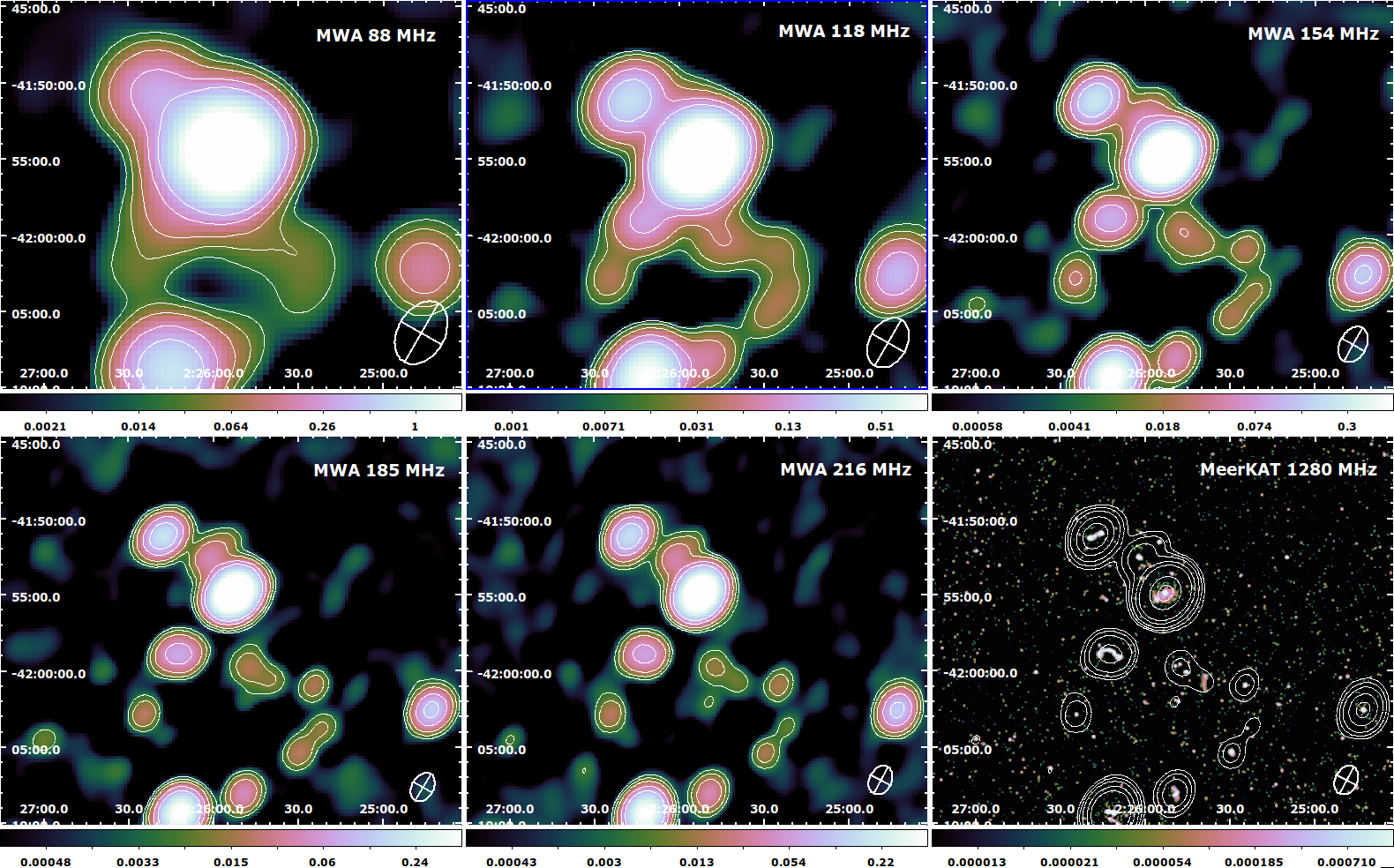}
    \caption{MWA-2 robust $+1.0$ images of A3016-A3017 across five frequencies. White contours show 5$\sigma$, 10$\sigma$, 20$\sigma$, 50$\sigma$, and 100$\sigma$ intensity level. MeerKAT full-resolution image at 1280~MHz is also presented with overlaid MWA 216~MHz contours. }
    \label{fig:mwa5}
\end{figure}

\section{Convolved MeerKAT 1280~MHz images}

\begin{figure}[h]
    \centering
    \includegraphics[width=0.45\textwidth]{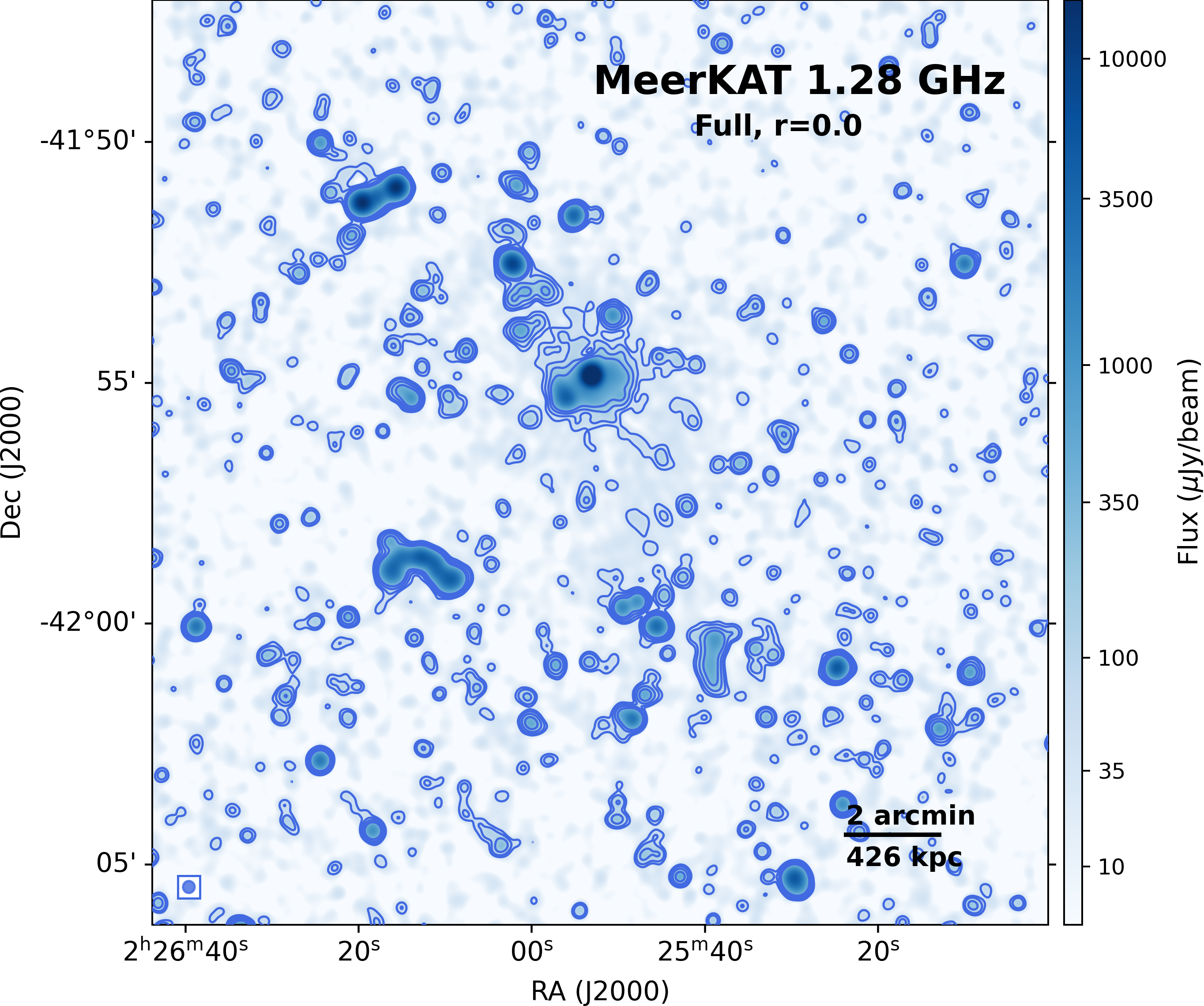}
    \includegraphics[width=0.45\textwidth]{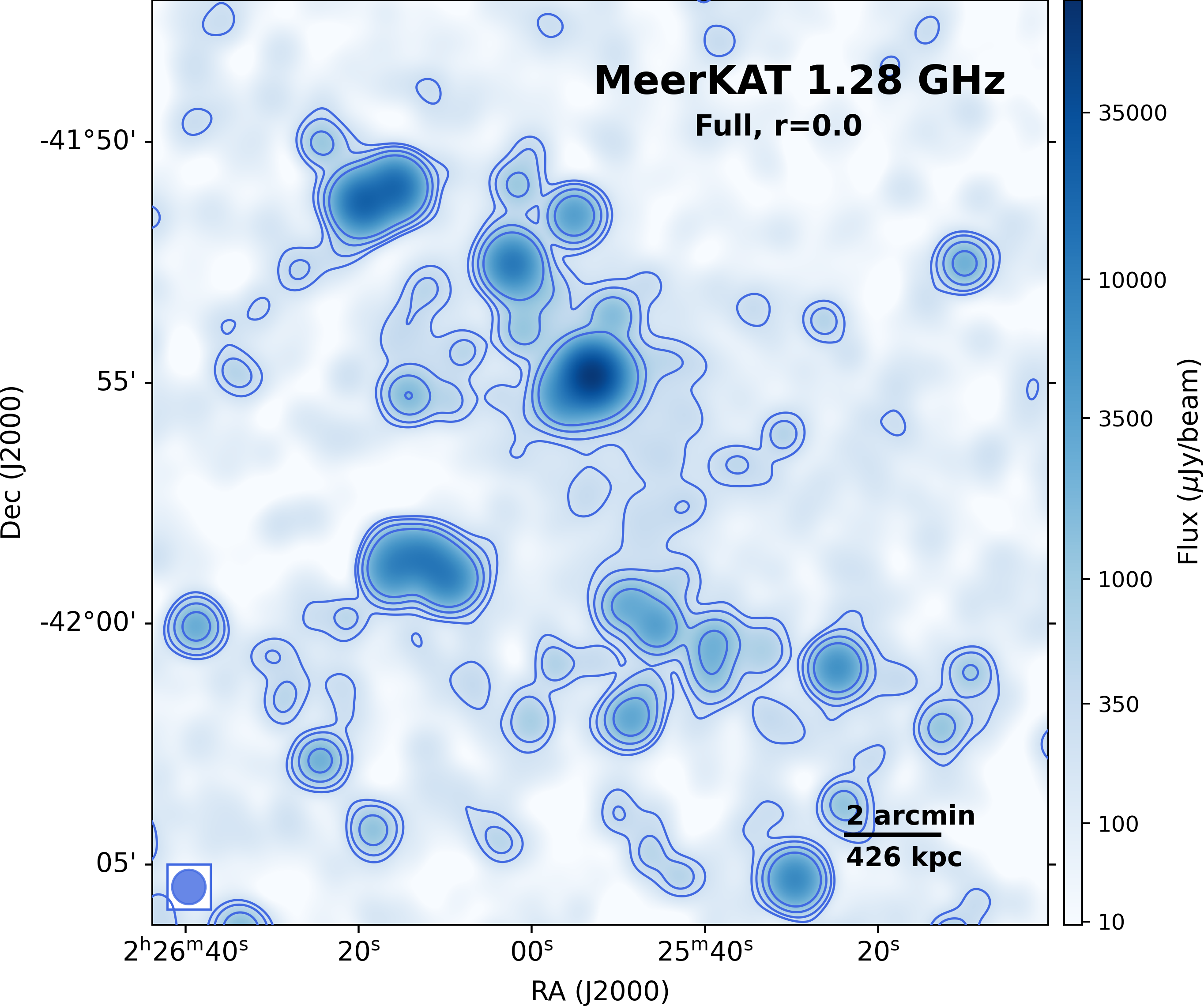}
    \caption{MeerKAT 1280~MHz images convolved to $15\arcsec$ (left) and $43\arcsec$ (right) resolutions from comparing with those compact-source subtracted images. Contours show 3$\sigma$, 5$\sigma$, 10$\sigma$, and 20$\sigma$ intensity level, with $\sigma$ values of $1.55\times 10^{-5}$~$\rm Jy~beam^{-1}$ and $8.12\times 10^{-5}$~$\rm Jy~beam^{-1}$, respectively. }
    \label{fig:convlved}
\end{figure}

\twocolumn

\end{appendix}

\end{document}